\newcommand{\avg}[1]{\left< #1 \right>} 

\documentclass{aastex}
\usepackage{amsbsy}
\usepackage{amsfonts}
\usepackage{amssymb}
\usepackage{longtable}
\usepackage{lscape}

\usepackage{graphicx}
\usepackage{graphics}
\usepackage{dcolumn}
\usepackage{bm}
\begin{document}


\title{ Properties of a Polar Coronal Hole \\ During the Solar Minimum in 2007}

\author{M. Hahn\altaffilmark{1}, P. Bryans\altaffilmark{2,3}, E. Landi\altaffilmark{4,5}, M. P. Miralles\altaffilmark{6}, and D. W. Savin\altaffilmark{1}}
\altaffiltext{1}{Columbia Astrophysics Laboratory, Columbia University, MC 5247, 550 West 120th Street, New York, NY 10027 USA}
\altaffiltext{2}{George Mason University, 4400 University Drive, Fairfax, Virginia 22030}
\altaffiltext{3}{Present address: NASA Goddard Space Flight Center, 8800 Greenbelt Road, Greenbelt, Maryland 20771. }
\altaffiltext{4}{US Naval Research Laboratory, Space Science Division, 4555 Overlook Avenue, SW, Code 76000A, Washington, DC 20375, USA}
\altaffiltext{5}{Present address: Department of Atmospheric, Oceanic and Space Sciences, University of Michigan, Ann Arbor, MI 48109}
\altaffiltext{6}{Harvard-Smithsonian Center for Astrophysics, 60 Garden Street, Cambridge, MA 02138, USA}

\date{\today}
\begin{abstract}

	We report measurements of a polar coronal hole during the recent solar minimum using the Extreme Ultraviolet Imaging Spectrometer on \emph{Hinode}. Five observations are analyzed that span the polar coronal hole from the central meridian to the boundary with the quiet Sun corona. We study the observations above the solar limb in the height range of $1.03 - 1.20$~$R_{\sun}$. The electron temperature $T_{\mathrm{e}}$ and emission measure EM are found using the Geometric mean Emission Measure (GEM) method. The EM derived from the elements Fe, Si, S, and Al are compared in order to measure relative coronal-to-photospheric abundance enhancement factors. We also studied the ion temperature $T_{\mathrm{i}}$ and the non-thermal velocity $v_{\mathrm{nt}}$ using the line profiles. All these measurements are compared to polar coronal hole observations from the previous (1996-1997) solar minimum and to model predictions for relative abundances. There are many similarities in the physical properties of the polar coronal holes between the two minima at these low heights. We find that $T_{\mathrm{e}}$, $n_{\mathrm{e}}$, and $T_{\mathrm{i}}$ are comparable in both minima. $T_{\mathrm{e}}$ shows a comparable gradient with height. Both minima show a decreasing $T_{\mathrm{i}}$ with increasing charge-to-mass ratio $q/M$. A previously observed upturn of $T_{\mathrm{i}}$ for ions above $q/M > 0.25$ was not found here. We also compared relative coronal-to-photospheric elemental abundance enhancement factors for a number of elements. These ratios were $\sim 1$ for both the low first ionization potential (FIP) elements Si and Al and the marginally high FIP element S relative to the low FIP element Fe, as is expected based on earlier observations and models for a polar coronal hole. These results are consistent with no FIP effect in a polar coronal hole. 
	
\end{abstract}

\maketitle
	
\section{Introduction} 
	
	A coronal hole is a relatively cool, low-density, and open field-line region of the solar atmosphere. During solar minimum, large coronal holes are found at the Sun's polar regions as was seen in the 1996-1997 and 2007-2009 solar minima. Coronal holes are also known source regions of the fast solar wind \citep{Krieger:SolPhys:1973, Zirker:Book:1977}, which is considered the basic equilibrium form of the solar wind \citep{Bame:JGR:1977}. 
	
	There are a number of differences between the recent 2007-2009 solar minimum and the previous minimum in 1996-1997. The polar magnetic field during the 2007-2009 solar minimum was about 40\% weaker and the polar coronal hole area was about 20\% smaller than during the 1996-1997 minimum \citep{Wang:ApJ:2009}. Measurements of the solar wind from the \emph{Ulysses} spacecraft show that the fast solar wind during the 2007-2009 minimum was 3\% slower and 17\% less dense compared to the previous one \citep{Issautier:GRL:2008,McComas:GRL:2008}. Those measurements also show that the solar wind power decreased by 25\% relative to the 1996-1997 solar minimum. 
	
	 In this paper we investigate the physical properties of a polar coronal hole in 2007 at low heights and compare them to those of the previous minimum. We examine spectral data from the Extreme Ultraviolet Imaging Spectrometer (EIS) instrument \citep{Korendyke:ApplOpt:2006, Culhane:SolPhys:2007} onboard the \emph{Hinode} satellite \citep{Kosugi:SolPhys:2007}. From the EIS spectral data we measure the electron density $n_{\mathrm{e}}$, electron temperature $T_{\mathrm{e}}$, emission measure EM, and ion temperature $T_{\mathrm{i}}$. We also looked for any ``FIP effect''. Elements with a low first ionization potential (FIP~$\lesssim 10$~eV) can undergo what has been dubbed as the FIP effect in which the elemental abundances of low FIP ions are enhanced in the corona relative to the photosphere. These ratios of coronal-to-photospheric abundances are called ``FIP factors''. We investigated the FIP effect by comparing ratios of FIP factors. All these results can be compared to similar measurements from extreme ultraviolet spectra collected during the 1996-1997 solar minimum with the Solar Ultraviolet Measurement of Emitted Radiation Spectrometer \citep[SUMER;][]{Wilhelm:SolPhys:1995} and the Coronal Diagnostic Spectrometer \citep[CDS;][]{Harrison:SolPhys:1995}, both on the Solar and Heliospheric Observatory (\emph{SOHO}) spacecraft \citep{Domingo:SolPhys:1995}.
	 	 
	The rest of this paper is organized as follows. The observations are described in Section~\ref{sec:obs}. Section~\ref{sec:emanal} reviews the emission measure analysis, which serves as one of our primary diagnostics. Data reduction, including line selection and fitting, are described in Section~\ref{sec:anal}. Results are presented in Section~\ref{sec:res}, followed by a discussion and summary in Sections~\ref{sec:dis}. Uncertainties are quoted throughout at an estimated $1\sigma$ statistical accuracy. 

\section{Observations}	\label{sec:obs}

	The observations were carried out with the EIS instrument onboard the \emph{Hinode} satellite. The spectrometer covers the wavelength ranges 171 - 211~\AA~and 245 - 291~\AA~and has a spectral resolution of 0.022~\AA~per pixel. There are $1^{\prime\prime}$, $2^{\prime\prime}$, $40^{\prime\prime}$, and $266^{\prime\prime}$ slit widths available, each observing a length of $512^{\prime\prime}$. The observations described here were all performed with the $2^{\prime\prime}$ slit. \citet{Brown:ApJS:2008} has measured the instrumental line width (FWHM) for the $1^{\prime\prime}$ slit to be 0.054~\AA~in the short wavelength band and 0.057~\AA~in the long wavelength band. The instrumental width for the $2^{\prime\prime}$ slit is estimated to be 0.007~\AA~greater, based on comparisons between observations of the same quiet Sun region with both the $1^{\prime\prime}$ and $2^{\prime\prime}$ slits \citep{Young:wiki:2009}.
	
\begin{figure}[h]
\begin{center}
\unitlength1cm
\begin{minipage}{8.0cm}
\resizebox{8.0cm}{!}{\includegraphics{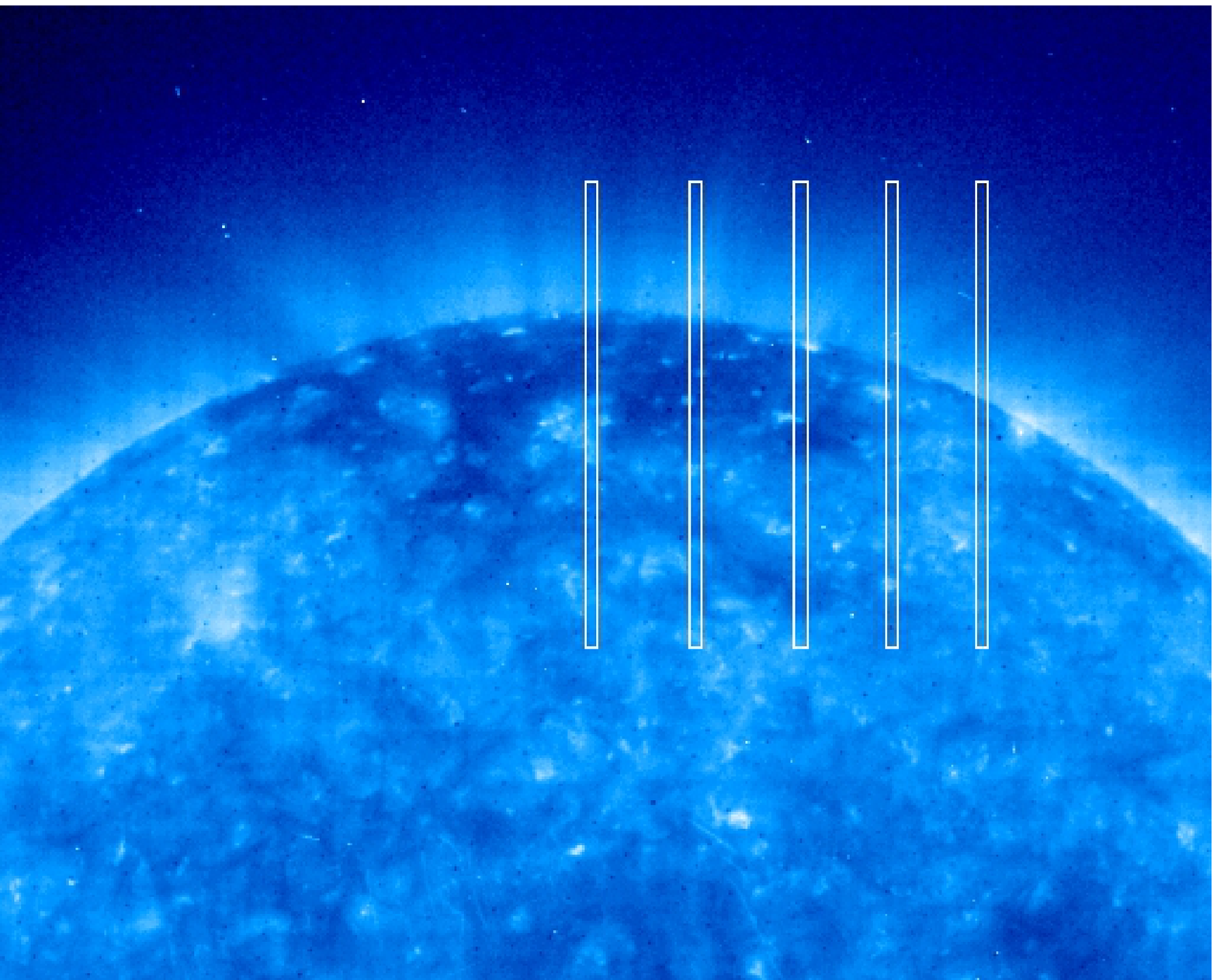}}
\centering 171 \AA 
\end{minipage}
\begin{minipage}{8.0cm}
\resizebox{8.0cm}{!}{\includegraphics{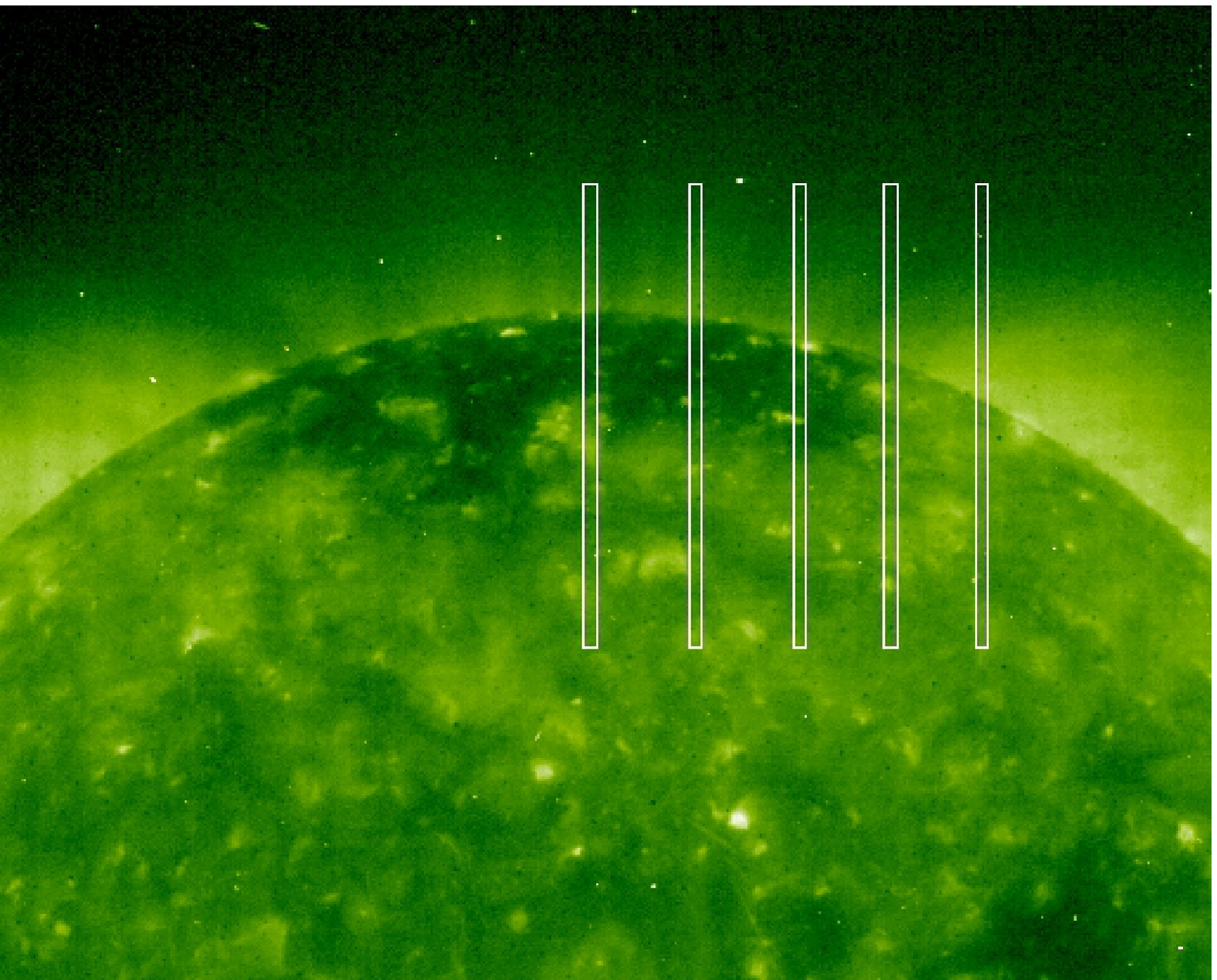}}
\centering 195 \AA
\end{minipage}
\end{center}
\caption{\label{fig:slitpicture} \emph{Solar and Heliospheric Observatory (SOHO)} Extreme ultraviolet Imaging Telescope \citep[EIT,][]{Delaboudiniere:SolPhys:1995} observations at 171~\AA~and 195~\AA, corresponding to lines from Fe IX/X ($\log T_{\mathrm{e}} = 5.9$) and Fe XII ($\log T_{\mathrm{e}} = 6.2$) lines, respectively. Here $T_{\mathrm{e}}$ is measured in units of K. The $14^{\prime\prime} \times 512^{\prime\prime}$ windows showing the EIS observations are outlined. Only the portion of the observations above the solar limb was used in the analysis. The observations are labelled in the text by their horizontal position relative to the center of the Sun. The line-of-sight of the observation centered at $-7^{\prime\prime}$ is over the central meridian and from left to right the observations are centered at $-7^{\prime\prime}$, $108^{\prime\prime}$, $223^{\prime\prime}$, $324^{\prime\prime}$, and $423^{\prime\prime}$.} 
\end{figure}

	For this study we analyze five observations made on 2007 November 16 at times 06:11, 06:49, 07:26, 09:40, and 10:50 UT. In each observation the 2$^{\prime\prime} \times$ 512$^{\prime\prime}$ slit was rastered across 7 positions in the horizontal direction giving a 14$^{\prime\prime} \times$ 512$^{\prime\prime}$ field of view. The centers of the horizontal scans were at $X= -7^{\prime\prime}$, 108$^{\prime\prime}$, 223$^{\prime\prime}$, 324$^{\prime\prime}$, and 423$^{\prime\prime}$ respectively, relative to the center of the Sun (Figure~\ref{fig:slitpicture}). Throughout this paper the observations are labelled according to their solar-$X$ location. In the vertical direction  all the observations were centered at $Y=879^{\prime\prime}$ relative to the center of the Sun. The line of sight of the portion of $-7^{\prime\prime}$ observation above the limb in the north polar coronal hole is roughly over the central meridian. The observations centered at $108^{\prime\prime}$ and $223^{\prime\prime}$ are further from the central meridian, but the above-limb slit portions are within the polar coronal hole. The $324^{\prime\prime}$ and $423^{\prime\prime}$ observations cover the quiet Sun corona at low radii and appear to cross into the polar coronal hole in the upper portion of the observation. 
		
		For the analysis here we were only interested in the coronal plasma above the limb. We therefore chose only pixels above the limb for analysis. As a result the portion of the EIS observations used in this study included each of the 7 horizontal pixels (spanning $14^{\prime\prime}$ at $2^{\prime\prime}$ per pixel) and above the limb about 150-250 vertical pixels ($1^{\prime\prime}$ per pixel). The spatial part of the observations below the limb was used to estimate the scattered light intensity as described below, but was not considered for the rest of the analysis. 
	
\section{Emission Measure Analysis}\label{sec:emanal}
	
	We use the Geometric mean Emission Measure (GEM) method of \citet{Bryans:ApJ:2009} to determine $T_{\mathrm{e}}$ and EM based on the intensity of observed spectral lines. The flux of a spectral line emitted by a transition from level $j$ to level $i$ of charge state $q$ for element $\mathrm X$ observed from distance $d$ is given by	
\begin{equation}
I_{ji} = \frac{1}{4\pi d^{2}} \int G(T_{\mathrm{e}},n_{\mathrm e}) n_{\mathrm e}^{2} \mathrm{d}V. 
\label{eq:intensityintegral}
\end{equation}
$G(T_{\mathrm{e}}, n_{\mathrm e})$ is the contribution function defined as 
\begin{equation}
G(T_{\mathrm{e}}, n_{\mathrm e}) \equiv \frac{n_{j}(\mathrm X^{+q})}{n(\mathrm X^{+q})}\frac{n(\mathrm X^{+q})}{n(\mathrm X)}\frac{n(\mathrm X)}{n(\mathrm H)}\frac{n(\mathrm H)}{n_{\mathrm e}}\frac{A_{ji}}{n_{\mathrm e}}.
\label{eq:Gdef}
\end{equation}
Here $n_{j}(\mathrm X^{+q})/n(\mathrm X^{+q})$ is the relative population of the upper level $j$ for ion $\mathrm{X}^{+q}$, $n(\mathrm X^{+q})/n(\mathrm X)$ is the relative abundance of charge state $q$ for element $\mathrm{X}$, $n(\mathrm X)/n(\mathrm H)$ is the abundance of $\mathrm{X}$ relative to hydrogen, $n(\mathrm{H})/n_{\mathrm e}$ is the hydrogen abundance relative to free electrons, and $A_{ji}$ is the Einstein A-rate. 
	
	For an isothermal plasma of uniform density Eq.~(\ref{eq:intensityintegral}) simplifies to 
\begin{equation}
I_{ji} = \frac{1}{4\pi d^{2}} G(T_{\mathrm{e}}, n_{\mathrm e}) \mathrm{EM},
\label{eq:simpleI}
\end{equation}
where EM is the emission measure defined as 
\begin{equation}
\mathrm{EM} = \int n_{\mathrm e}^2 \mathrm{d}V.
\label{eq:EMdef}
\end{equation}
The above equations can be solved for the EM as a function of $T_{\mathrm{e}}$ and $n_{\mathrm{e}}$ giving
\begin{equation}
\mathrm{EM} = 4 \pi d^{2} \frac{I_{ji}}{G(T_{\mathrm{e}},n_{\mathrm{e}})}.
\label{eq:EMofT}
\end{equation}
The observed data provide the intensities of the various spectral lines. The function $G(T_{\mathrm{e}}, n_{\mathrm{e}})$ is calculated from atomic data. For most lines $G(T_{\mathrm{e}}, n_{\mathrm{e}})$ is only a very weak function of $n_{\mathrm{e}}$. The EM as a function of temperature can then be plotted for each emission line using equation~(\ref{eq:EMofT}) where $n_{\mathrm{e}}$ is fixed. The density used in the analysis was estimated using line intensity ratios (described in Section~\ref{subsec:ne}). According to the definition of equation~(\ref{eq:EMdef}) the emission measure depends only on the electron density and the volume of the emitting plasma. The assumptions made here imply that these properties should be the same for all the ions in the volume, therefore an analysis using the intensity of different emission lines originating from the same volume should yield the same EM. The temperature is determined by finding the point $(\mathrm{EM},T_{\mathrm{e}})$ which best satisfies the condition that the EM should be the same for each line. When the $\mathrm{EM}(T_{\mathrm{e}})$ curves are plotted, this condition is met at the point where the curves intersect. 

	 The GEM method is a particular way of extracting the average values of EM and $T_{\mathrm{e}}$ using the intersection of the $\mathrm{EM}(T_{\mathrm{e}})$ curves for the observed lines \citep{Bryans:ApJ:2009}. Because $\mathrm{EM}(T_{\mathrm{e}})$ varies rapidly, the GEM approach uses the mean of the logarithm (the geometric mean) rather than the arithmetic mean. When multiple intersections occur for a given pair of EM curves we eliminate the less physically probable crossings by choosing the intersection closest to the average crossing point for all the EM curves. We also eliminate crossings between EM curves from the same ion as such crossings are most likely due to observational issues or uncertainties in the atomic data. Finally, we omit intersection points from temperatures where $n(\mathrm{X}^{+q})/n(\mathrm{X})$ is below $1\%$ since under that condition the collisional ionization equilibrium calculations are known to be less reliable \citep{Bryans:ApJS:2006}.
	
	 $G(T_{\mathrm{e}}, n_{\mathrm e})$ is calculated using the \textsc{chianti} atomic database \citep{Dere:AA:1997,Dere:AA:2009} and the collisional ionization equilibrium data from \citet{Bryans:ApJ:2009}. The elemental abundances $n(\mathrm X)/n(\mathrm H)$ are assumed to be photospheric, but relative coronal abundances can be inferred with further analysis as discussed in Section~\ref{sec:res}. The photospheric abundances are taken from \citet{Asplund:ARAA:2009}. 
		
\section{Analysis of Line Intensities} \label{sec:anal}

	The data were reduced using the standard EIS preparation routines to remove the dark current, cosmic ray spikes, and warm pixels. For warm pixels we followed the procedure of \citet{Young:WarmPix:2009}. The output of the EIS preparation routines converts measured counts to physical units ($\mathrm{erg}\,\mathrm{cm^{-2}\,s^{-1}\,sr^{-1}\,\AA^{-1}}$) with an uncertainty based on Poisson counting statistics, dark current, and removal of the warm pixels. 
	
	Because emission from the polar coronal hole is faint, to improve statistics we collapsed these reduced data into bins consisting of 10 vertical pixels and all 7 horizontal pixels. The averaging routine used to collapse the data was careful to ignore bad pixels flagged by the preparation routines. 

	Lines were chosen for analysis based on the identifications of \citet{Young:PASJ:2007}, \citet{Brown:ApJS:2008}, \citet{Young:ApJ:2009}, and \citet{Landi:ApJ:2009b}. Table~\ref{table:linelist} lists all lines used. Note that from here on all wavelengths are given in units of~\AA. We chose lines that were either free from significant blends, or where blends were separated in wavelength so that the component peaks could clearly be distinguished, or where the blend contribution could be subtracted out based on a measurement of a third unblended line known to have a fixed intensity ratio with one of the components of the blended line. We also used the Fe~\textsc{x} $\lambda$257.26, Fe~\textsc{xii} $\lambda$195.12, and Fe~\textsc{xiii} $\lambda$203.82 lines, which are inseparable self blends. In each of these cases, since the blend comes from the same ion the total intensity can be used in the GEM analysis by summing the contribution functions $G(T_{\mathrm{e}}, n_{\mathrm{e}})$ for each component without introducing additional uncertainties from differences in elemental or charge state abundances. 
	
	For each observation we used 25 lines from Fe \textsc{viii}-\textsc{xiii} and 8 lines from Si \textsc{vii}, \textsc{ix}, and \textsc{x}. In the 423$^{\prime\prime}$ observation, our furthest outward slit position, the electron temperature was greater than at other positions and some Fe \textsc{xiv} lines were strong enough to be used in the analysis. For elements other than iron and silicon there were very few lines in these observations that were both reasonably strong and unblended. We found 2 S~\textsc{viii} lines, 2 S~\textsc{x} lines, 1 Al~\textsc{viii} line, 3 Al~\textsc{ix} lines, and 3 Mg~\textsc{vii} lines. All of the Al lines were very weak and could only be analyzed for the lowest few spatial bins above the limb. Both S~\textsc{viii} lines are blended with Fe~\textsc{xi} lines. The Fe~\textsc{xi} $\lambda$198.55 intensity, which blends with the S~\textsc{viii} $\lambda 198.55$ line, can be estimated by measuring the intensity of the Fe~\textsc{xi} $\lambda$189.13 line. The theoretical intensity ratio is Fe~\textsc{xi} $\lambda 198.55 / \lambda 189.13 = 0.84$. Similarly, the blended contribution from Fe~\textsc{xi} $\lambda 202.63$ to the S~\textsc{viii} $\lambda 202.61$ line can be subtracted by measuring the Fe~\textsc{xi} $\lambda 188.23$ line intensity where the theoretical ratio Fe~\textsc{xi} $\lambda 202.63/\lambda 188.23 = 0.017$. Lastly, the Mg~\textsc{vii} $\lambda$278.40 line is a blend with Si~\textsc{vii} $\lambda 278.45$, which has a theoretical intensity 0.323 times that of the Si~\textsc{vii} $\lambda$275.36 line. In all three cases the pair of lines in the ratio originate from the same upper level and hence the intensity ratio is determined solely by the branching ratio for transitions to the lower levels.

\begin{figure}
\begin{center}
\unitlength1cm
\begin{minipage}{8.0cm}
\resizebox{8.0cm}{!}{\includegraphics{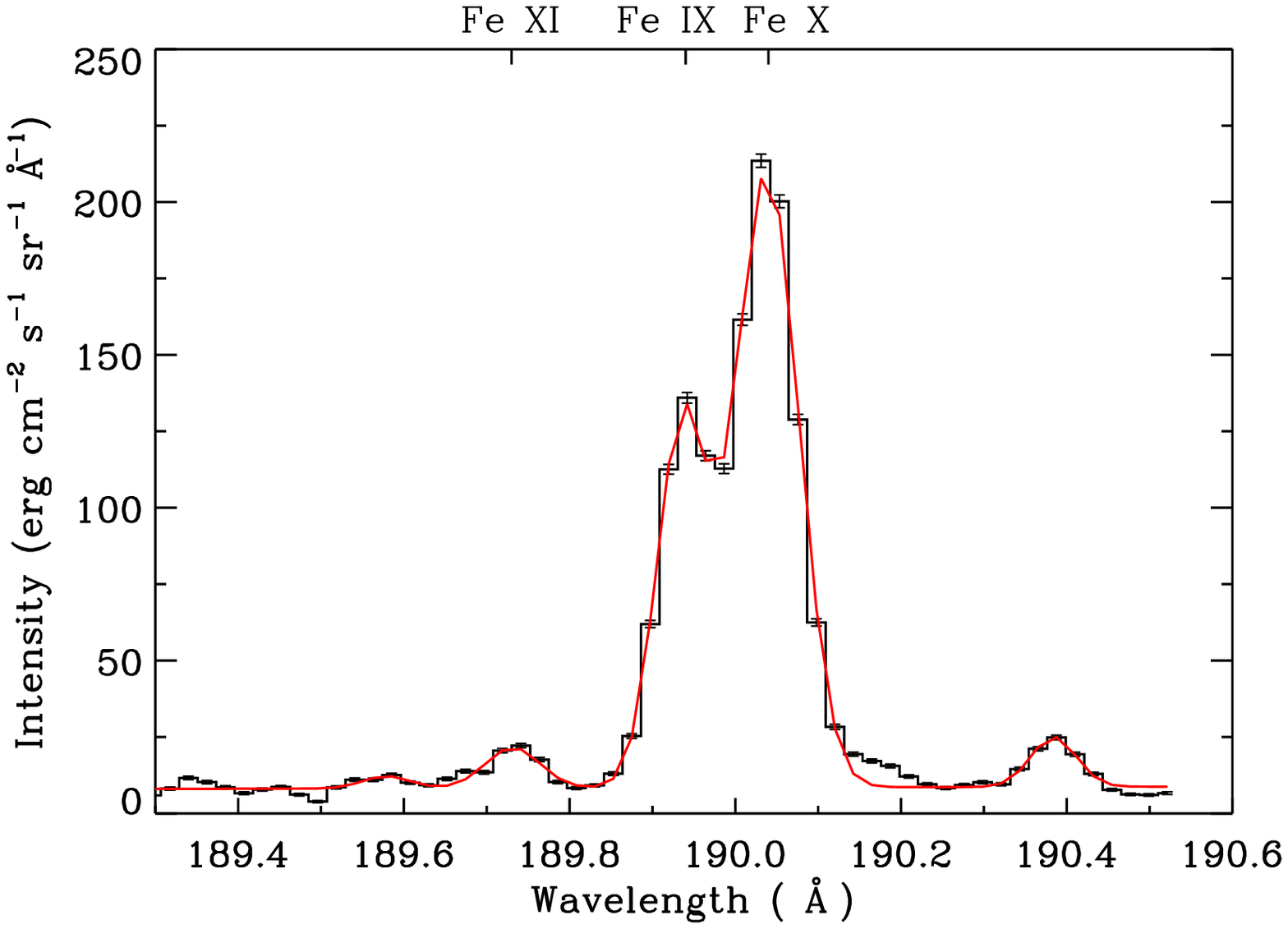}}
\end{minipage}
\begin{minipage}{8.0cm}
\resizebox{8.0cm}{!}{\includegraphics{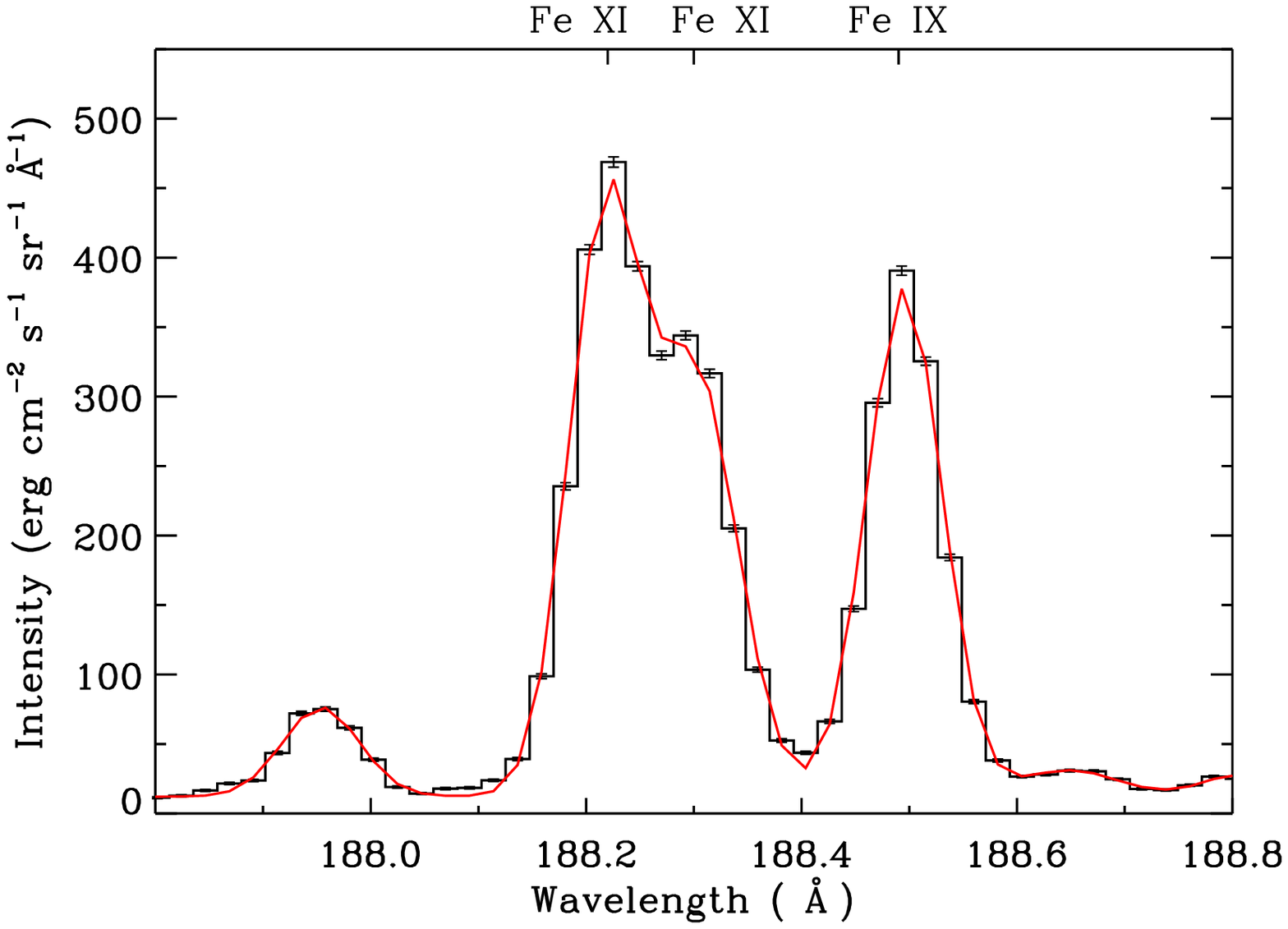}}
\end{minipage}
\end{center}
\caption{\label{fig:examplefits} Examples of emission line fits using multiple Gaussians. Shown are Fe~\textsc{xi} 189.72~\AA, Fe~\textsc{ix} 189.94~\AA, and Fe~\textsc{x} 190.04~\AA~lines (left) and  Fe~\textsc{xi} 188.22~\AA, Fe~\textsc{xi} 188.30~\AA, and Fe~\textsc{ix} 188.49~\AA~lines (right), which were used in the analysis. Both figures are for a height R $\approx 1.08$~$R_{\sun}$ in the $223^{\prime\prime}$ observation. These figures represent two of the most complex fits. In most cases lines were unblended or were more widely spaced than in the examples shown here.}
\end{figure}

	The intensities and widths of the emission lines were determined by fitting a sum of Gaussian functions plus a linear background to a small wavelength range around each line. The Gaussian fits used the \emph{eis\_auto\_fit\_gen} routine from the EIS analysis software. When fitting multiple lines over the same wavelength interval, constraints were included as needed in the fitting to keep the position of the line centroid close to the expected wavelengths. These constraints typically allowed less than $0.1$~\AA~freedom in the centroid position, corresponding to a line of sight velocity $\lesssim 150$~$\mathrm{km\,s^{-1}}$. This allowed Doppler shifts up to this velocity to be detectable. However, we found that centroid positions were very close to the expected wavelengths. Comparing fits using slightly different sets of constraints confirmed that systematic errors arising from the fitting method did not affect the GEM analysis results. Examples of two of the most complex fits are shown in Figure~\ref{fig:examplefits}, but the majority of fits were of more widely separated lines.

	Line intensities beyond the solar limb consist of coronal line emission plus a component from the solar disk scattered within the instrument. Recently the scattered light intensity was measured during a partial solar eclipse. It was found that pixels pointed at the eclipsed portion of the solar disk read 2\% of the intensity of the pixels observed in the uneclipsed portion of the solar disk \citep{Ugarte:ScatLight:2010}. This scattered light component was removed in the current analysis by measuring the intensity of each line for several bins below the limb, which was nearly constant, and subtracting 2\% of that average intensity from the total intensity of the above-limb pixels used in the analysis. For some lines the scattered light subtraction brought the intensity in the highest spatial bins to essentially zero.
		
\section{Results} \label{sec:res}
	
\subsection{Density} \label{subsec:ne}
	
	We used the ratio of emission lines from the same ion to measure the electron density. Our observations are of a low density region close to the limb and under these circumstances photo-excitation and stimulated emission from photospheric black body radiation can be important processes \citep{Young:ApJ:2003}. Hence, it was necessary to include these effects in the density analysis. Not accounting for these effects gave a density that is too large by about 50\%.
	
\begin{figure}
\includegraphics[width=0.9\textwidth]{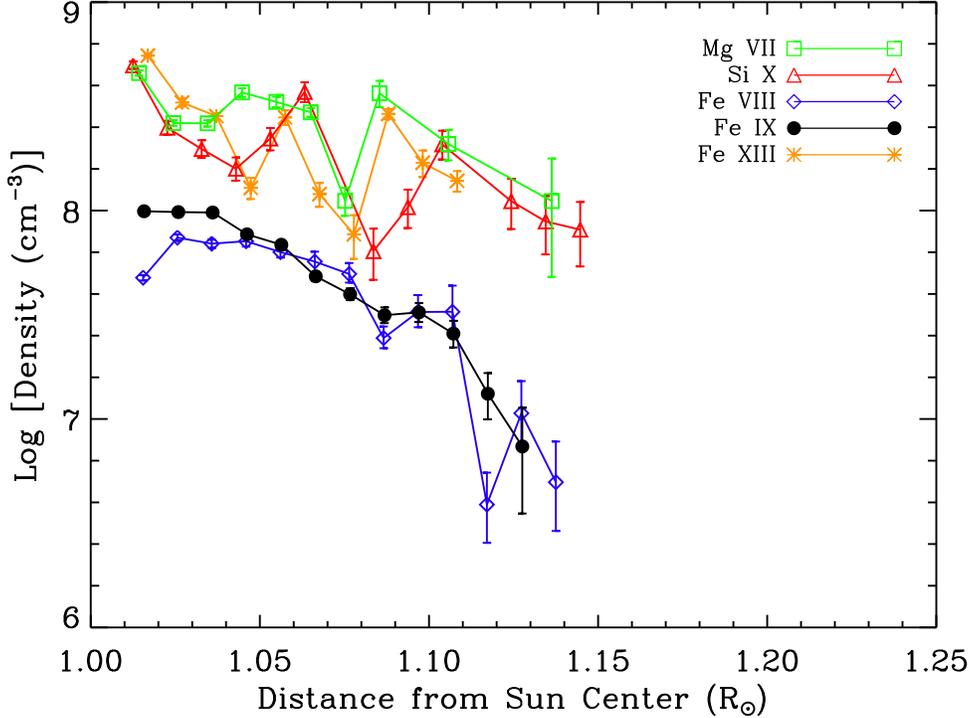}
\caption{\label{fig:DensityCompare} Density profile for the $223^{\prime\prime}$ observation using several different density diagnostic line ratios. The Fe~\textsc{viii} $\lambda 186.60/ \lambda 185.21$ and Fe~\textsc{ix} $\lambda 189.94/\lambda 188.49$ ratios yield a density in the range expected for polar coronal holes. The Si~\textsc{x} $\lambda 258.38/\lambda 261.04$, Fe~\textsc{xiii} $(\lambda 203.83 + \lambda 203.80)/\lambda202.04$, and Mg~\textsc{vii} $\lambda 280.75/276.15$ diagnostics imply a density a factor of $\sim 2$ larger. The discrepancy for the Si~\textsc{x} and Fe~\textsc{xiii} diagnostics is likely an observational effect caused by the presence of a small amount of hotter quiet Sun corona along the line-of-sight that affects the higher temperature Si~\textsc{x} and Fe~\textsc{xiii} lines but not the cooler Fe~\textsc{viii} and \textsc{ix} lines. However, Mg~\textsc{vii} is abundant at typical coronal hole temperatures and the reason for the discrepancy between the Mg~\textsc{vii} line and the Fe~\textsc{viii} and \textsc{ix} lines is unknown. The error bars show only the statistical uncertainty from the fit. Overall uncertainties can be inferred from the scatter in the density profiles.}
\end{figure}

	In the polar coronal hole observations between $-7^{\prime\prime}$ to $324^{\prime\prime}$ the Fe~\textsc{viii} $\lambda 186.60 / \lambda 185.21$ and Fe~\textsc{ix} $\lambda 189.94 / \lambda 188.49$ intensity ratios show that the density falls from $\approx 8 \times 10^{7}$~$\mathrm{cm^{-3}}$ to $\approx 1 \times 10^{7}$~$\mathrm{cm^{-3}}$ between $1.05$~$R_{\sun}$ and $1.15$~$R_{\sun}$ (Figure~\ref{fig:DensityCompare}). The density inferred from these lines is about the same as what was found in polar coronal hole measurements from the previous solar minimum at similar heights \citep{Banerjee:AA:1998, Wilhelm:ApJ:1998, Fludra:JGR:1999, Landi:ApJ:2008}. 
	
	 The  Si~\textsc{x}~$\lambda 258.38 / \lambda 261.04$, Fe~\textsc{xiii}~$(\lambda 203.82 + \lambda 203.79)/\lambda 202.04$, and Mg~\textsc{vii}~$\lambda 280.75 / 276.15$ ratios, however, yield a density that is roughly a factor of 2 larger at $1.05$~$R_{\sun}$ and fall off slower with height compared to the Fe~\textsc{viii} and Fe~\textsc{ix} density diagnostics. This apparent discrepancy for the Si~\textsc{x} and Fe~\textsc{xiii} diagnostics is probably due to the presence of quiet Sun coronal plasma along the line-of-sight. For example, assuming the Fe~\textsc{viii} and \textsc{ix} lines give the actual coronal hole density and assuming typical quiet Sun coronal temperature and densities of $\log T_{\mathrm{e}} = 6.15$ and $n_{\mathrm{e}} = 2 \times 10^{8}$~$\mathrm{cm^{-3}}$ at 1.05~$R_{\sun}$, we estimate that the quiet Sun emission measure needs to be only a few percent of the coronal hole emission measure in order to produce the inferred densities. This is because the high formation temperatures of the Si~\textsc{x} and Fe~\textsc{xiii} lines causes the intervening quiet Sun corona to contribute significantly to the observed intensity. The Fe~\textsc{viii} and Fe~\textsc{ix} ions have a lower temperature of formation and are largely insensitive to quiet Sun corona, making them more reliable density diagnostics in the relatively cool polar coronal hole. An observational effect like this cannot explain the high density given by the Mg~\textsc{vii} lines since that ion is abundant at typical coronal hole temperatures. We note that discrepancy between the density derived from the Fe~\textsc{viii} and Mg~\textsc{vii} lines has been observed previously \citep{Young:ApJ:2009b}.	
	
	Figure~\ref{fig:slitpicture} shows that the observation at $423^{\prime\prime}$ appears to observe the quiet Sun corona at low heights. The observation also has a significantly higher temperature, as will be discussed in Section~\ref{subsec:te}. Consequently, the Si~\textsc{x} and Fe~\textsc{xiii} intensity ratios are the best density diagnostics to use in this case. The density in the $423^{\prime\prime}$ observation, based on these lines, was $2 \times 10^{8}$~$\mathrm{cm^{-3}}$ at $1.05$~$R_{\sun}$ and dropped to about $5 \times 10^{7}$~$\mathrm{cm}^{-3}$ at 1.15~$R_{\sun}$. This is in agreement with quiet Sun density measurements, such as those of \citet{Feldman:ApJ:1999}. 
	
\subsection{Electron Temperature}\label{subsec:te}

	To measure the temperature as a function of height we performed the GEM analysis using the selected iron and silicon for each observation (see Table~\ref{table:linelist}). These elements had several charge states and a large number of lines available so we consider them to be more reliable indicators of the temperature than sulfur, aluminum, or magnesium for which there were few observed lines. The GEM analysis was performed for each of the 7~pixel~$\times$~10~pixel spatial bins that were above the limb brightening, usually at distances from the center of the Sun $R \gtrsim 1.03$~$R_{\sun}$. The density parameter for $G(T_{\mathrm{e}}, n_{\mathrm e})$ versus height was set using $n_{\mathrm e}$ from Fe~\textsc{viii}~and~\textsc{ix} for the $-7^{\prime\prime}$ to $324^{\prime\prime}$ observations and using Si~\textsc{x} and Fe~\textsc{xiii} for the $423^{\prime\prime}$ observation. We also performed the analysis with input densities that were larger and smaller by an order of magnitude and found that while a small number of lines are density sensitive, the average EM and $T_{\mathrm{e}}$ are insensitive to the input density. 
	
\begin{figure}
\includegraphics[width=0.9\textwidth]{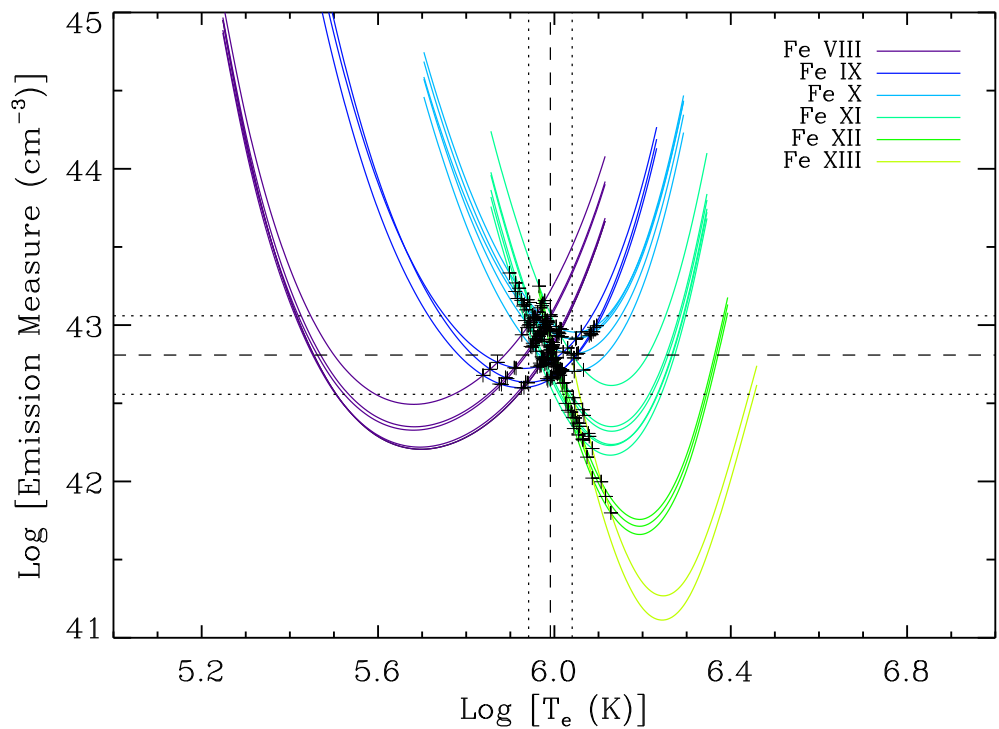}
\caption{\label{fig:GEMFe} Emission Measure $\mathrm{EM}(T_{\mathrm{e}})$ for the iron lines at a height of 1.06~$R_{\sun}$ in the $223^{\prime\prime}$ observation. The dashed lines show the mean values and the dotted lines show the 1$\sigma$ uncertainty ranges of $T_{\mathrm{e}}$ and EM. Here $\log T_{\mathrm{e}} = 5.99 \pm 0.05$ and $\log \mathrm{EM} = 42.81 \pm 0.25$. }
\end{figure}

\begin{figure}
\includegraphics[width=0.9\textwidth]{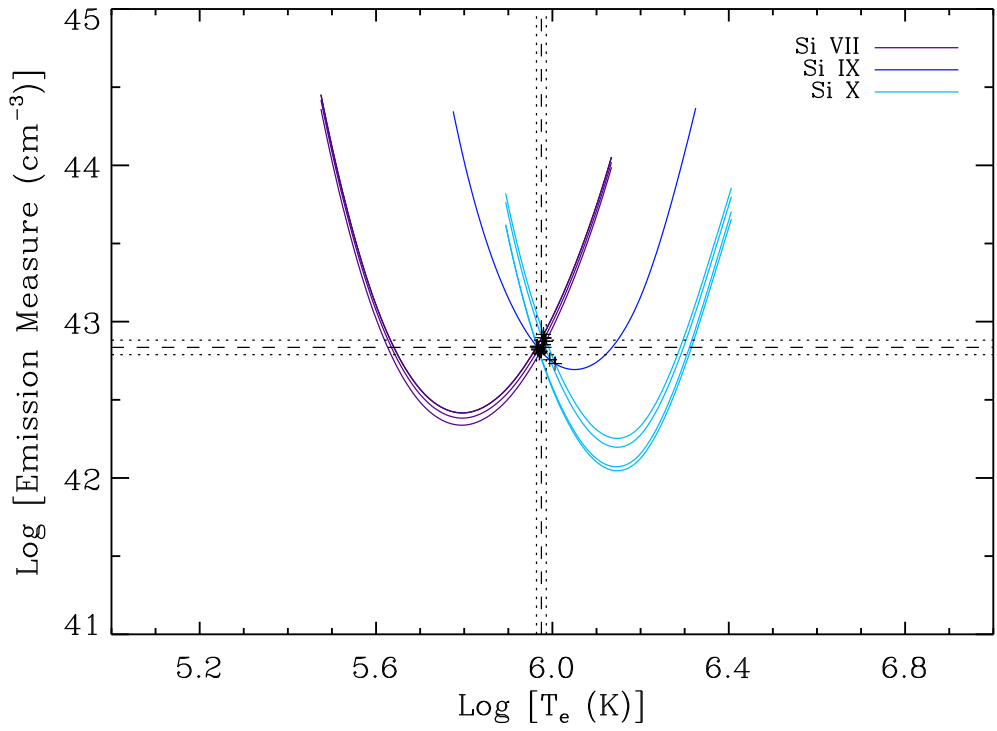}
\caption{\label{fig:GEMSi} Same as Figure~\ref{fig:GEMFe} but showing the $\mathrm{EM}(T_{\mathrm{e}})$ from the silicon lines. Here $\log T_{\mathrm{e}} = 5.97 \pm 0.01$ and $\log \mathrm{EM} = 42.84 \pm 0.05$.}
\end{figure}

	Examples of the EM analysis are shown in Figures~\ref{fig:GEMFe} and \ref{fig:GEMSi} for iron and silicon, respectively. The $\mathrm{EM}(T_{\mathrm{e}})$ curves for lines of each charge state are in reasonable agreement. Measuring temperature in units of K here and throughout the paper, the crossing points also appear in a tight cluster near $\log T_{\mathrm{e}} = 6$, which supports the idea that the plasma is approximately isothermal. The iron line analysis shows that there are some crossing points that extend to higher temperatures, which could indicate a higher temperature component or a systematic uncertainty. 
	
\begin{figure}
\includegraphics[width=0.9\textwidth]{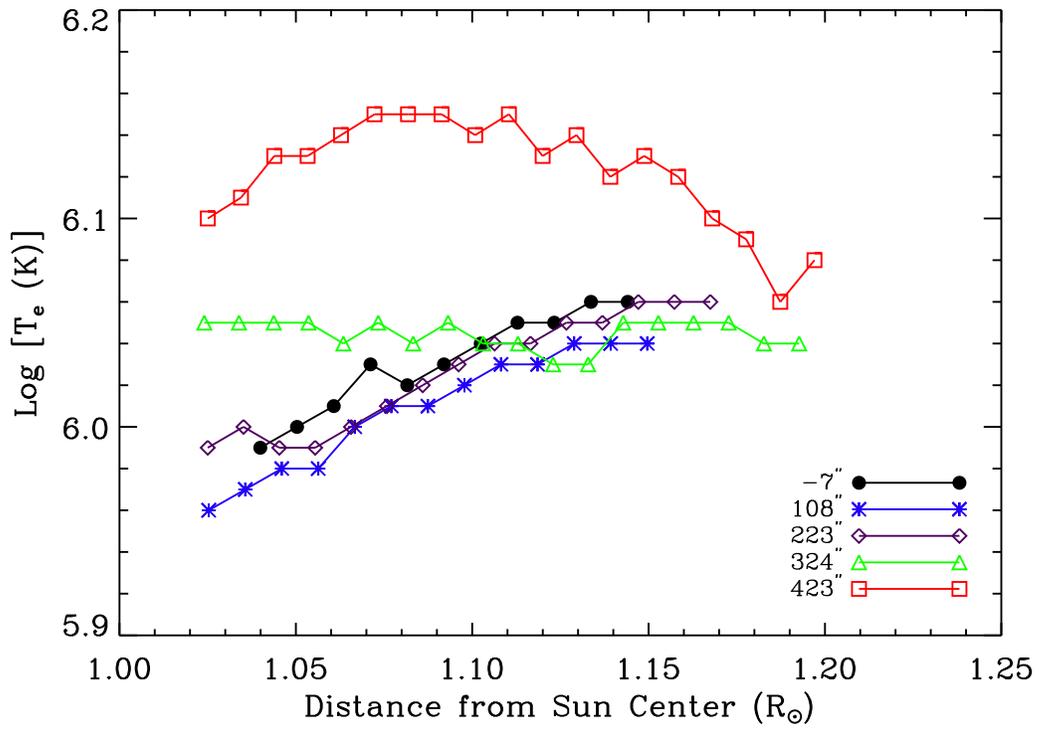}
\caption{\label{fig:TeFe} Electron temperature profiles for the various observations from the GEM analysis using iron emission lines. The $1\sigma$ uncertainties in $\log T_{\mathrm{e}}$ are typically $\pm (0.05 - 0.07)$. }
\end{figure}

\begin{figure}
\includegraphics[width=0.9\textwidth]{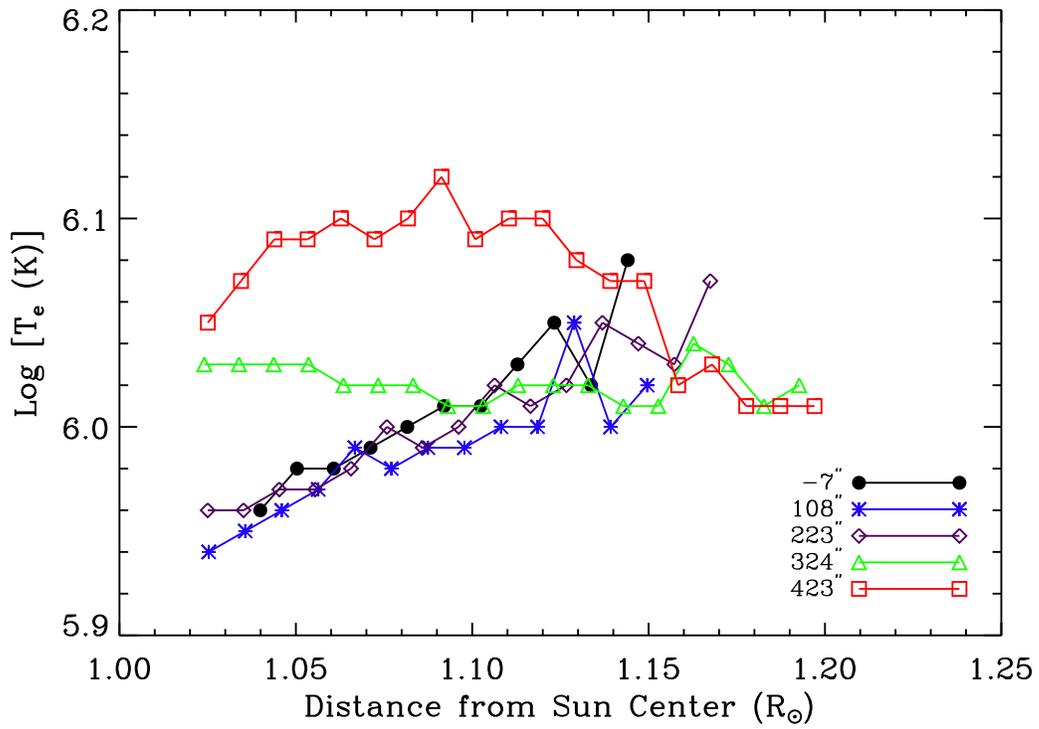}
\caption{\label{fig:TeSi} Same as Figure~\ref{fig:TeFe}, but using the silicon lines. The $1\sigma$ uncertainties in $\log T_{\mathrm{e}}$ here are typically $\pm (0.02-0.04)$. }
\end{figure}

\begin{figure}
\includegraphics[width=0.9\textwidth]{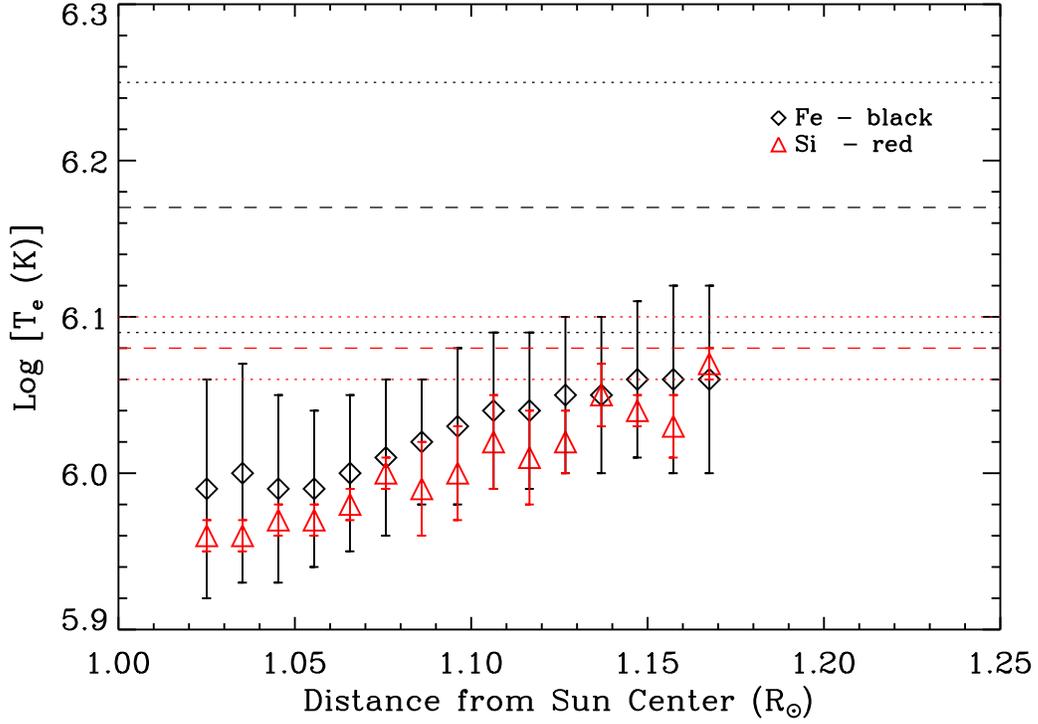}
\caption{\label{fig:Te072624} Temperatures from the GEM analysis of iron and silicon lines for the 223$^{\prime\prime}$ observation. The black and red dashed horizontal lines show the $T_{\mathrm{e}}$ inferred from the scale height analysis of iron and silicon, respectively. The dotted lines illustrate the associated uncertainties on the scale height $T_{\mathrm{e}}$. The scale height temperature results are summarized in Table~\ref{table:Tscale}.}
\end{figure}

	Figures~\ref{fig:TeFe} and \ref{fig:TeSi} show the temperature profiles from iron and silicon, respectively. The temperature measurements using either the Fe or the Si lines were in agreement within the uncertainties, but the iron $T_{\mathrm{e}}$ is systematically greater than that from silicon (e.g.,~Figure~\ref{fig:Te072624}). The iron lines come from a larger number of charge states having temperatures of maximum abundance in the range $\log T_{\mathrm{e}} = 5.6 - 6.3$. In contrast, the silicon lines come from only three charge states with temperatures of maximum abundance in the range $\log T_{\mathrm{e}} = 5.8 - 6.1$. The result is that there is a larger spread in the $\mathrm{EM}(T_{\mathrm{e}})$ crossing points for Fe than for Si and a correspondingly greater uncertainty. The large number of iron lines with a maximum abundance $\log T_{\mathrm{e}} > 6.1$ appear to systematically increase the average temperature derived from the iron lines compared to the silicon lines. Fe~\textsc{xii}-\textsc{xiv} have maximum abundances at $\log T_{\mathrm{e}} > 6.2$. Removing these lines from the analysis removes the systematic temperature difference.
	
	The three observations in the polar coronal hole at $-7^{\prime\prime}$, 108$^{\prime\prime}$, and 223$^{\prime\prime}$ all showed similar temperature profiles, increasing from about $\log T_{\mathrm{e}} \approx 5.95$~near 1.03~$R_{\sun}$ to $\log T_{\mathrm{e}} \approx 6.02$ near 1.15~$R_{\sun}$ (Figures~\ref{fig:TeFe} and \ref{fig:TeSi}). The temperature increases steadily with a gradient $\Delta \log T_{\mathrm{e}} / \Delta R \approx 0.75$~$R_{\sun}^{-1}$ up to about 1.12~$R_{\sun}$ and then levels off above that height. 
	
	These inferred temperatures are very similar to measurements from the previous solar minimum \citep{Doschek:ApJ:1998, Feldman:ApJ:1998, Fludra:JGR:1999, Doschek:ApJ:2001, Landi:ApJ:2008}. Earlier polar coronal hole observations also showed that the temperature increases with height. We find a temperature gradient very similar to that of \citet{Landi:ApJ:2008}. A possible systematic cause of the apparent temperature increase is the presence of surrounding hot quiet Sun coronal plasma along the line-of-sight. Since the scale height of the higher temperature plasma is larger, the intensity from intervening hot, quiet Sun plasma would decrease with height more slowly than the colder polar coronal hole plasma. This would be observed as a temperature gradient \citep{Feldman:PoP:2008}. \citet{Doschek:ApJ:2001} performed a DEM analysis for a similar observation taken during the previous solar minimum and also found a temperature gradient. Their DEM showed a peak in the EM distribution at $\log T_{\mathrm{e}} \approx 5.95$, but the ratio of high temperature to low temperature DEM increased with height.
		
	In the 324$^{\prime\prime}$ observation, $T_{\mathrm{e}}$ was roughly constant over the height range of the observation with $\log T_{\mathrm{e}} \approx 6.03 - 6.05$. Below about 1.10~$R_{\sun}$, $T_{\mathrm{e}}$ is clearly greater than in the $-7^{\prime\prime}$ to 223$^{\prime\prime}$ polar coronal hole observations, but above 1.10~$R_{\sun}$ the temperature matches that of the $-7^{\prime\prime}$ to $223^{\prime\prime}$ observations. Since 324$^{\prime\prime}$ is near the edge of the polar coronal hole this behavior could be caused by quiet Sun corona being observed simultaneously with the polar coronal hole at low heights, leading to the calculation of an intermediate temperature in our analysis. 
	
	The 423$^{\prime\prime}$ observation appears to cover the boundary between the quiet Sun corona and the polar coronal hole. At low heights the observation falls in the quiet Sun corona and the temperature increases from $\log T_{\mathrm{e}} \approx 6.08$ at $1.03$~$R_{\sun}$ to $\log T_{\mathrm{e}} \approx 6.15$ at $1.10$~$R_{\sun}$. The temperature is then constant up to $\approx 1.15$~$R_{\sun}$ where it begins to decline. Figure~\ref{fig:slitpicture} shows that $1.15$~$R_{\sun}$ is close to the intensity transition between the quiet Sun corona and the polar coronal hole. Above 1.18~$R_{\sun}$ the temperature has dropped to about the level of the other polar coronal hole observations. The quiet Sun temperature agrees very well with previous observations, such as those reported by \citet{Feldman:ApJ:1999} who found $\log T_{\mathrm{e}} \approx 6.11$.  
	
\subsection{Scale Height Electron Temperature}\label{subsec:scaleheight}	

\begin{figure}
\includegraphics[width=0.9\textwidth]{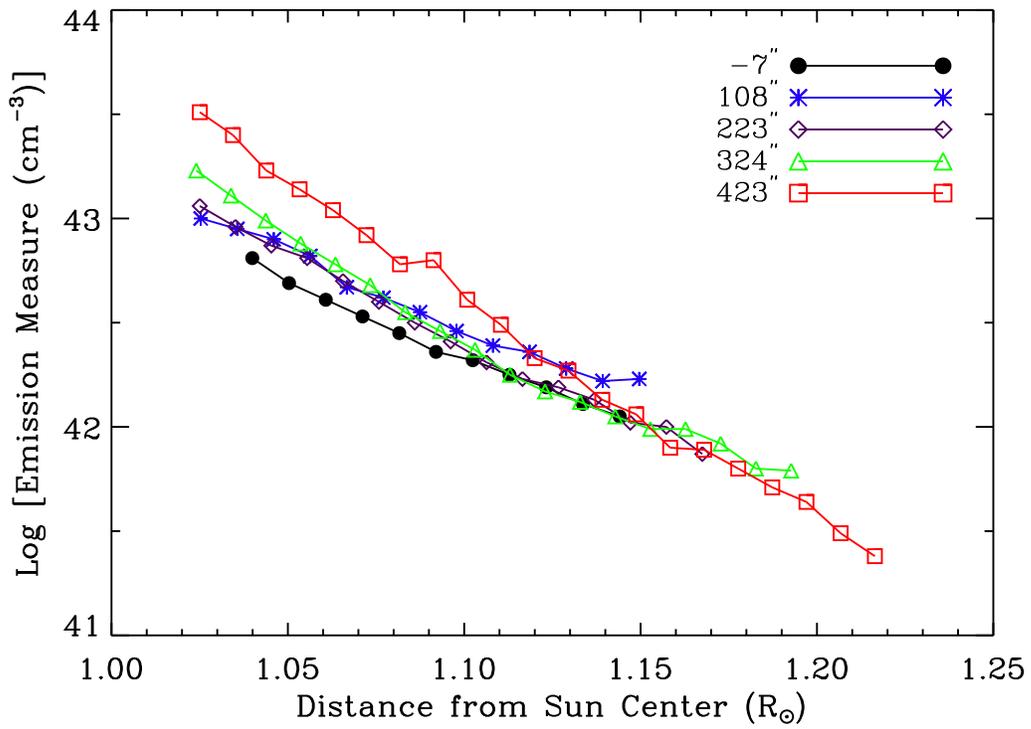}
\caption{\label{fig:EMem} Emission measure obtained using the GEM method for the iron lines in each observation. The $1\sigma$ uncertainties on $\log \mathrm{EM}$ are typically $\pm (0.2 - 0.3)$.}
\end{figure}

\begin{figure}
\includegraphics[width=0.9\textwidth]{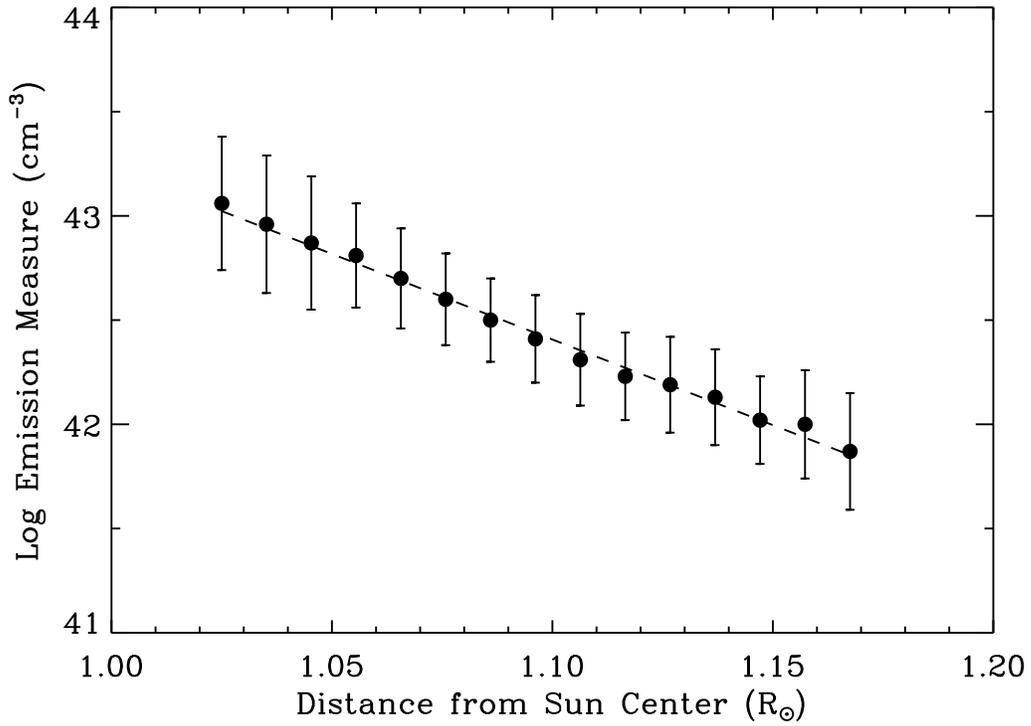}
\caption{\label{fig:EMscale} The emission measure and $1\sigma$ uncertainties obtained from the analysis of the iron lines in the 223$^{\prime\prime}$ observation. The linear fit was used to measure the scale height and estimate $T_{\mathrm{e}}$. In this case $\log T_{\mathrm{e}} = 6.17 \pm 0.08$. }
\end{figure}

  In every observation the EM as a function of height decreases exponentially (Figure~\ref{fig:EMem}). In each case the emission measures and slopes of EM versus height from Fe and Si were in agreement to within the uncertainties. The slope of the EM curves is related to the scale height and can be used to estimate the temperature. For an isothermal plasma in hydrostatic equilibrium the density is proportional to
\begin{equation}
n_{\mathrm e} \propto \exp(-mgR /k_{\mathrm B}T_{\mathrm{e}}),
\label{eq:nehydro}
\end{equation}
where $R$ is radial distance, $g$ is the surface gravity of the Sun, $k_{\mathrm{B}}$ is the Boltzmann constant, and $m$ is the mean particle mass, about 0.61 $m_{\mathrm{H}}$ with $m_{\mathrm{H}}$ the hydrogen mass. The emission measure is a volume integral of the square of the density. If the spatial bins are small, then the density does not vary greatly across the field of view and the volume integral reduces to an integral along the line-of-sight. Since the radial distance across the observation is small, here about $0.15$~$R_{\sun}$, the length through the plasma along the line-of-sight will be approximately constant over this height. The EM is then related to the radius by
\begin{equation}
\mathrm{EM} \propto \exp(- 2mgR/k_{\mathrm B}T_{\mathrm{e}}) .
\label{eq:emhydro}
\end{equation}
Using this, the electron temperature can be estimated from the slope of a linear fit to $\ln(\mathrm{EM})$ versus radial distance (Figure~\ref{fig:EMscale}). The temperatures for each observation using this method are given in Table~\ref{table:Tscale}. Performing the analysis over shorter sub-intervals did not reveal any systematic change in the slope of $\mathrm{EM}(R)$, so we report the results for a linear fit over the full above-limb distance. In the polar coronal hole observations $-7^{\prime\prime}$, 108$^{\prime\prime}$, and $223^{\prime\prime}$ the iron and silicon scale height temperatures are greater than the temperatures measured by the emission measure analysis (e.g., Figures~\ref{fig:TeFe}, \ref{fig:TeSi}, and \ref{fig:Te072624}). However, the present scale height temperature measurement is closer to $T_{\mathrm{e}}$ measured using EM analysis methods than has been found in the past. A similar scale height analysis of a polar coronal hole during the 1996-1997 solar minimum found a larger difference between the two methods with $\log T_{\mathrm{e}} \approx 6.5$ from the scale height compared to $\log T_{\mathrm{e}} \approx 6.0$ from an emission measure analysis \citep{Landi:ApJ:2008}. In the $324^{\prime\prime}$ observation $T_{\mathrm{e}}$ derived from the scale height is about the same as that derived from the GEM analysis and in $423^{\prime\prime}$ observation the scale height $T_{\mathrm{e}}$ is smaller than $T_{\mathrm{e}}$ from the GEM analysis. 

\subsection{FIP factors}\label{subsec:fip}

	Elemental abundances in the corona may differ from their photospheric values. It is commonly observed that elements with a first ionization potential (FIP) below the hydrogen Lyman~$\alpha$ energy of 10.199~eV are enhanced in the corona relative to the photosphere. This is referred to as the FIP effect and the amount of the enhancement is called the FIP factor. Based on previous observations and theoretical models, coronal holes are expected to have a small FIP effect with FIP factors of $\approx 1-2$ compared to $\approx 3-4$ in the quiet Sun corona \citep{Feldman:ApJ:1998, Zurbuchen:GPRL:2002, Bryans:ApJ:2009, Laming:ApJ:2009}.

	The EM for two elements can be used to determine ratios of FIP factors. We have calculated the EM using photospheric values for the relative abundance $n(\mathrm{X})/n(\mathrm{H})$. Coronal abundances can be enhanced by the FIP factor, which is defined as
\begin{equation}
f_{\mathrm{X}} \equiv \frac{\left[n(\mathrm{X})/n(\mathrm{H})\right]_{\mathrm {corona}}}{\left[n(\mathrm{X})/n(\mathrm{H})\right]_{\mathrm {photosphere}}}.
\label{deffsubx}
\end{equation}
One can convert $G(T_{\mathrm{e}}, n_{\mathrm{e}})$ based on photospheric abundances into a $G(T_{\mathrm{e}}, n_{\mathrm{e}})$ based on coronal abundances by multiplying the photospheric abundance based $G(T_{\mathrm{e}},n_{\mathrm{e}})$ by $f_{\mathrm{X}}$. Since $\mathrm{EM} \propto 1/G(T_{\mathrm{e}}, n_{\mathrm{e}})$ this means $\mathrm{EM_{c}} = \mathrm{EM_{p}}/f_{\mathrm{X}}$, where the subscripts denote quantities inferred using coronal (c) versus photospheric (p) abundances. In the logarithm this relation is
\begin{equation}
\log(\mathrm{EM_{\mathrm{c}}})=\log(\mathrm{EM_{p}}) - \log(f_{\mathrm{X}}).
\label{eq:trueem}
\end{equation}
If all the emission lines come from the same volume in the corona and the correct coronal abundances are used, then $\mathrm{EM_{c}}$ should be independent of the element whose emission lines are used to calculate it. We can therefore determine the ratio of FIP enhancement factors for two elements $f_{\mathrm{X}}$ and $f_{\mathrm{Y}}$ by 
\begin{equation}
\log(f_{\mathrm{Y}}/f_{\mathrm{X}}) = \log[\mathrm{EM_{p}}(\mathrm Y)] - \log[\mathrm{EM_{p}}(\mathrm X)].
\label{eq:fipfactorratio}
\end{equation}

\subsubsection{Iron and Silicon}\label{subsubsec:fesi}

\begin{figure}
\includegraphics[width=0.9\textwidth]{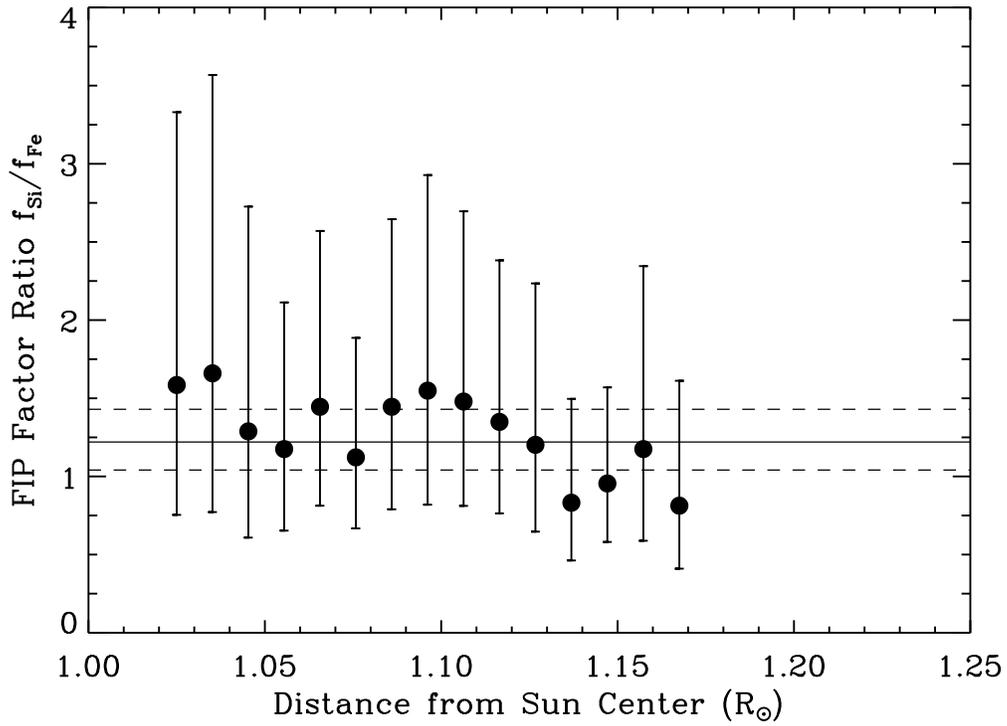}
\caption{\label{fig:EMratio} Ratio of the FIP factors $f_{\mathrm{Si}}/f_{\mathrm{Fe}}$ plotted as a function of height for the $223^{\prime\prime}$ observation. The weighted mean and $1\sigma$ uncertainty are indicated by the solid and dashed horizontal lines, respectively. The values for each observation are tabulated in Table~\ref{table:fipratio}}
\end{figure}

	 Both Fe and Si are low FIP elements with FIPs of 7.902~eV and 8.152~eV, respectively. Since both are low FIP elements, they are expected to have similar FIP factors with small differences due to differing ionization and recombination rates. None of the observations showed a significant trend in $f_{\mathrm{Si}}/f_{\mathrm{Fe}}$ as a function of height. An example showing the 223$^{\prime\prime}$ observation is presented in Figure~\ref{fig:EMratio}. The uncertainty weighted mean values $\avg{f_{\mathrm{Si}}/f_{\mathrm{Fe}}}$ for each observation are tabulated in Table~\ref{table:fipratio}. The mean is calculated in the logarithm where the uncertainties are symmetric. For all polar coronal hole observations here, the ratio was consistent with 1. In the quiet Sun corona in the $423^{\prime\prime}$ observation the ratio is sightly greater than 1. 
	 
	 The present results are in agreement with previous polar coronal hole observations, which find $f_{\mathrm{Si}}/f_{\mathrm{Fe}} \approx 1$ \citep{Feldman:ApJ:1998, Zurbuchen:GPRL:2002}. Observations of the quiet Sun corona find that the $f_{\mathrm{Si}}/f_{\mathrm{Fe}}$ ratio remains close to 1 \citep{Feldman:ApJ:1998, Zurbuchen:GPRL:2002, Bryans:ApJ:2009}. \citet{Laming:ApJ:2009} presents a model in which the FIP effect is caused by the ponderomotive force generated by Alfv\'{e}n waves propagating through the chromosphere and being transmitted or reflected within coronal loops. This model predicts $f_{\mathrm{Si}}/f_{\mathrm{Fe}} \approx 1$ in a coronal hole and $f_{\mathrm{Si}}/f_{\mathrm{Fe}} = 1.0 - 1.5$ in quiet Sun regions. Both values are consistent with our results and previous observations, although the uncertainties in the current measurements limit our ability to make a more detailed comparison with theory.

\subsubsection{Sulfur}\label{subsubsec:su}
	
\begin{figure}
\includegraphics[width=0.9\textwidth]{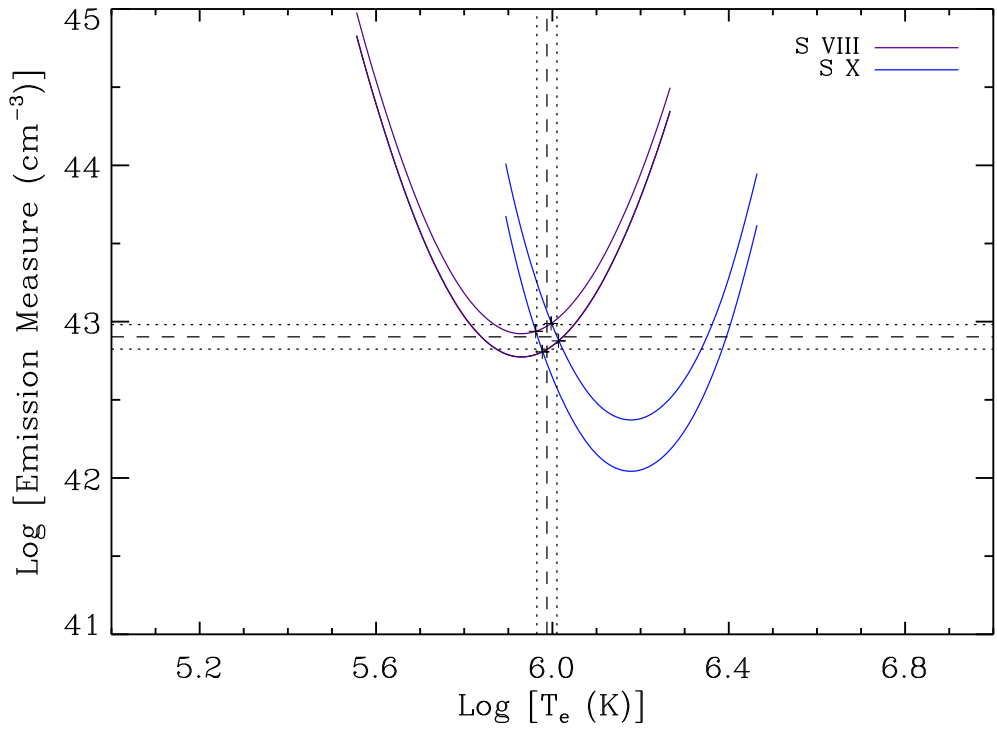}
\caption{\label{fig:Scurves} Same as Figure~\ref{fig:GEMFe} but showing $\mathrm{EM}(T_{\mathrm{e}})$ from the sulfur lines. Here $\log T_{\mathrm{e}} = 5.99 \pm 0.02$ and $\log \mathrm{EM} = 42.90 \pm 0.08$. }
\end{figure}

		Figure~\ref{fig:Scurves} shows the GEM analysis for the four sulfur lines we observed. The temperatures derived from these lines are in agreement with the temperatures from the iron and silicon lines in every observation. The emission measures were used to calculate $f_{\mathrm{S}}/f_{\mathrm{Fe}}$. The mean values for $\avg{f_{\mathrm{S}}/f_{\mathrm{Fe}}}$ are presented in Table~\ref{table:fipratio}. None of the observations showed a trend in the ratio as a function of height. In most observations we found $\avg{f_{\mathrm{S}}/f_{\mathrm{Fe}}} \approx 0.91-1.39$. The upper part of the $423^{\prime\prime}$ observation, above the apparent $T_{\mathrm{e}}$ transition to cooler plasma, had a slightly smaller value, $\avg{f_{\mathrm{S}}/f_{\mathrm{Fe}}} \approx 0.82$, but this is within the error bars of the measurements for the other pointings.

	As mentioned earlier, there appears to be a small amount of quiet Sun corona along the line-of-sight. Since the S~\textsc{x} lines are more abundant at quiet Sun coronal temperatures and are therefore more sensitive to such plasmas, any quiet Sun coronal plasma will increase the observed intensity of the S~\textsc{x} lines relative to the S~\textsc{viii} lines. This increases both the EM and the $T_{\mathrm{e}}$ at which the crossing points occur in the GEM analysis and thereby increases the inferred $f_{\mathrm{S}}/f_{\mathrm{Fe}}$. Thus, the observed $f_{\mathrm{S}}/f_{\mathrm{Fe}}$ is actually an upper limit for the polar coronal hole measurement. 
		
	The first ionization potential of sulfur is 10.360~eV, making sulfur a marginally high FIP element. Sulfur is therefore expected to undergo little FIP enhancement relative to low FIP elements. In a coronal hole the FIP effect model of \citet{Laming:ApJ:2009} predicts that the FIP effect is weak and neither sulfur nor iron experience a significant enhancement, giving $f_{\mathrm{S}} \approx f_{\mathrm{Fe}} \approx 1$ and consequently $f_{\mathrm{S}}/f_{\mathrm{Fe}} \approx 1$. Our polar coronal hole observations find that $f_{\mathrm{S}}/f_{\mathrm{Fe}}$ is close to 1, which suggests no FIP effect was present in agreement with models. 

	Some polar coronal hole observations during the previous solar minimum showed $f_{\mathrm{S}}/f_{\mathrm{Fe}} < 1$ \citep{Feldman:ApJ:1998}. These results might be explained by the revisions to the photospheric elemental abundance data. If we use photospheric abundances from \citet{Feldman:PhysScr:2000} rather than \citet{Asplund:ARAA:2009} we find $f_{\mathrm{S}}/f_{\mathrm{Fe}} \sim 0.6 - 0.7$ in the polar coronal hole observation, which is similar to that found by \citet{Feldman:ApJ:1998}.  
				
	The FIP effect is expected to be stronger in the quiet Sun corona. Therefore it is expected that in the quiet Sun corona $f_{\mathrm{S}} / f_{\mathrm{Fe}} < 1$. The \citet{Laming:ApJ:2009} model predicts $f_{\mathrm{S}}/f_{\mathrm{Fe}} \approx 0.5 - 0.9$. Observations of solar equatorial regions \citep{Feldman:ApJ:1998,Bryans:ApJ:2009}, as well measurements of the slow solar wind \citep{Zurbuchen:GPRL:2002} were consistent with a FIP effect for sulfur. However, we do not observe a FIP effect near the polar coronal hole in the quiet Sun portion of the $423^{\prime\prime}$ observation where $\avg{f_{\mathrm{S}}/f_{\mathrm{Fe}}} = 1.12^{+0.18}_{-0.15}$. 
		
\subsubsection{Aluminum}\label{subsubsec:al}

\begin{figure}
\includegraphics[width=0.9\textwidth]{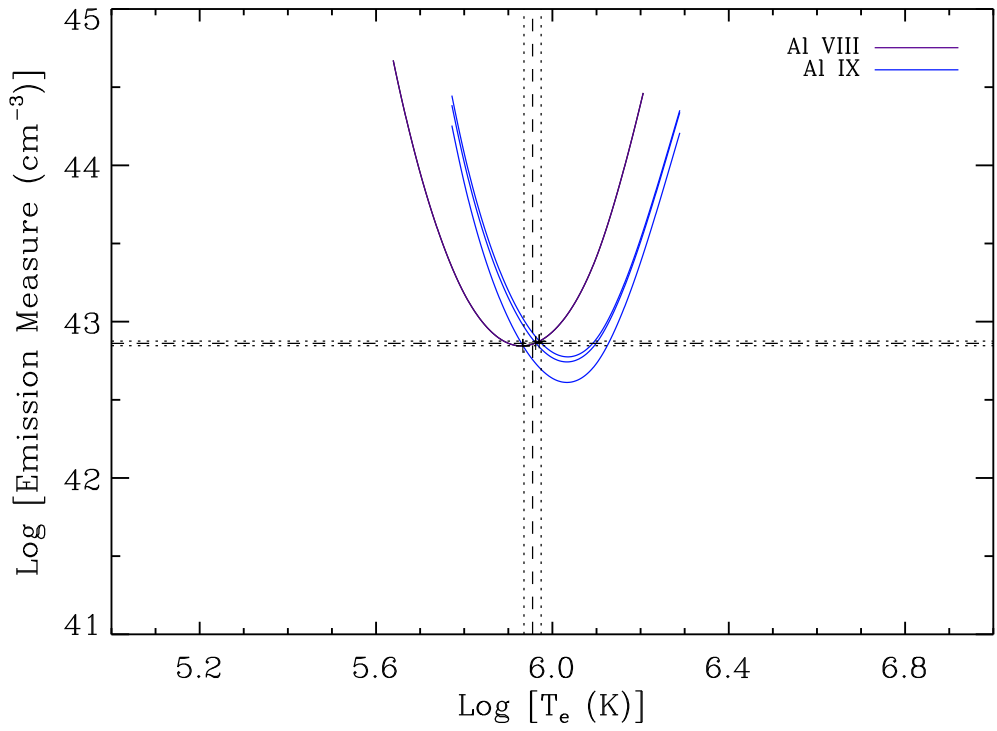}
\caption{\label{fig:Alcurves} Same as Figure~\ref{fig:GEMFe} but showing $\mathrm{EM}(T_{\mathrm{e}})$ from the aluminum lines. Here $\log T_{\mathrm{e}} = 5.96 \pm 0.02$ and $\log \mathrm{EM} = 42.86 \pm 0.01$.}
\end{figure}

	Figure~\ref{fig:Alcurves} shows our GEM analysis for aluminum lines. Only four Al lines were sufficiently intense that they could be used in our analysis and even then for only a few of the lowest spatial bins with $R \lesssim 1.10$~$R_{\sun}$. The lines come from only two charge states that give only three crossing points for the analysis. The small number of lines and crossing points means that our use of the standard deviation is not an accurate representation of the uncertainty. In those locations where the GEM analysis could be applied to the aluminum lines the temperature was similar to the results from other lines.
	
	Al is a low FIP element with an FIP of 5.986~eV. We calculated the relative abundance factors for Al compared to Fe for all observations. The values for $\avg{f_{\mathrm{Al}}/f_{\mathrm{Fe}}}$ are shown in Table~\ref{table:fipratio}. For the $423^{\prime\prime}$ observation the Al lines could only be measured in the lower (quiet Sun) part of the slit. In every observation $f_{\mathrm{Al}}/f_{\mathrm{Fe}}$ was not significantly different from 1. This is consistent with expectations since both Al and Fe are low FIP elements, which are expected to undergo similar FIP effects. For example, the \citet{Laming:ApJ:2009} model predicts $f_{\mathrm{Al}}/f_{\mathrm{Fe}} =1.0 - 1.1$ in a coronal hole and 1.2 in the quiet Sun corona. 
				
\subsubsection{Magnesium}\label{subsubsec:mg}

	  Only Mg~\textsc{vii} had strong unblended lines in these observations. Since no other charge states of Mg could be used to find crossing points, the GEM analysis method could not be applied. As a result we used a different method for estimating the EM. 
	  
\begin{figure}
\includegraphics[width=0.9\textwidth]{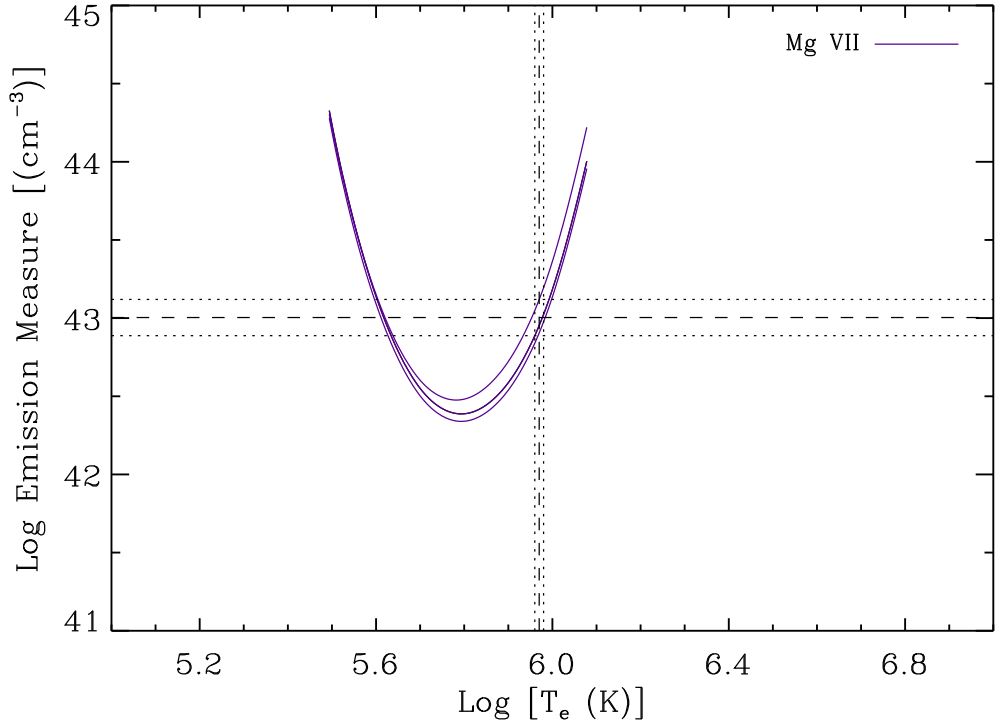}
\caption{\label{fig:MGcurves} $\mathrm{EM}(T_{\mathrm{e}})$ from the magnesium lines at 1.06~$R_{\sun}$ for the $223^{\prime\prime}$ observation. The EM was found using the temperature from the analysis of the silicon lines. Here assuming $\log T_{\mathrm{e}} = 5.97 \pm 0.01$ we found $\log \mathrm{EM} = 43.00 \pm 0.12$. }
\end{figure}

	  If the observed volume is isothermal we can measure the EM from the magnesium lines by using the temperature derived from the GEM analysis of a different element whose emission comes from the same volume. Here we use the $T_{\mathrm{e}}$ derived from the silicon lines which have a formation temperature closer to that for the magnesium lines than do the iron lines. We can then determine the EM for the magnesium lines by $\mathrm{EM} = \mathrm{EM}(T_{\mathrm{e, Si}})$, where by $T_{\mathrm{e,Si}}$ we mean the electron temperature derived from the silicon lines, not the temperature of the silicon ions themselves. In this case the uncertainty in the EM arises not only from the uncertainty in the measured intensity, which affects the magnitude of the $\mathrm{EM}$ vs. $T_{\mathrm{e}}$ curves and is always present, but also from the uncertainty on $T_{\mathrm{e,Si}}$. In order to account for this new source of uncertainty we performed a Monte-Carlo uncertainty analysis. We generated a Gaussian distribution of temperature values with the mean value $T_{\mathrm{e, Si}}$ and the standard deviation equal to the uncertainty. We then calculated $\mathrm{EM}(T_{\mathrm{e}})$ for each line at each temperature in the set. The mean and standard deviation of the resulting values give $\mathrm{EM}(T_{\mathrm{e,Si}})$ and its uncertainty. A plot of the $\mathrm{EM}(T_{\mathrm{e}})$ curves showing the result using this method is shown in figure~\ref{fig:MGcurves}. 
	  
	  The EM could not be accurately measured in the quiet Sun corona portion of the $423^{\prime\prime}$ observation. In that case the temperature is so high that the expected relative ion abundance of $n(\mathrm{Mg^{7+}})/n(\mathrm{Mg})$ is less than 1\% and the collisional ionization equilibrium calculations are known to be particularly uncertain \citep{Bryans:ApJS:2006, Bryans:ApJ:2009}. 
  
	  The $f_{\mathrm{Mg}}/f_{\mathrm{Fe}}$ ratio in the polar coronal hole observations was determined using the derived EM values. The mean $\avg{f_{\mathrm{Mg}}/f_{\mathrm{Fe}}}$ from each observation are presented in Table~\ref{table:fipratio}. The value of $f_{\mathrm{Mg}}/f_{\mathrm{Fe}}$ in the $-7^{\prime\prime}$ to $223^{\prime\prime}$ observations was about $1.4$. The ratio was somewhat larger in the $324^{\prime\prime}$ observation. This is likely due to interference from the nearby quiet Sun corona, which seems to be particularly strong in the $324^{\prime\prime}$ observation but could be a factor in the others as well. Intervening quiet sun corona would increase $T_{\mathrm{e,Si}}$. However, the Mg~\textsc{vii} lines are insensitive to the hotter quiet Sun corona. Thus, with $T_{\mathrm{e,Si}}$ larger than $T_{\mathrm{e}}$ in the volume where most Mg~\textsc{vii} emission actually occurs we would overstate the EM as well as the inferred $f_{\mathrm{Mg}}/f_{\mathrm{Fe}}$. 
	   	  	  
	  Magnesium is a low FIP element with FIP 7.646~eV. Both \citet{Feldman:ApJ:1998} and \citet{Zurbuchen:GPRL:2002} found that polar coronal hole and fast solar wind FIP factors for magnesium were only slightly larger than iron during the 1996-1997 solar minimum, with a value $f_{\mathrm{Mg}}/f_{\mathrm{Fe}} \approx 1.1$. The ponderomotive force model for the FIP effect predicts a ratio $\lesssim 1.2$ in a coronal hole \citep{Laming:ApJ:2009}. Thus, the value in the present measurement is larger than expected, but seems to agree within the uncertainty of the method.
	
\subsection{Ion Temperature}\label{subsec:Ti}

	The ion temperatures were estimated from the widths of the spectral lines. The width of a line is related to the ion temperature by \citep{Seely:ApJ:1997} 
\begin{equation}
\Delta\lambda_{\mathrm{FWHM}} = \left[ \Delta \lambda_{\mathrm{Inst}}^{2} + 4 \ln(2) \left( \frac{\lambda}{c} \right)^{2} \left(\frac{2 k_{\mathrm{B}}T_{\mathrm{i}}}{M} + v_{\mathrm{nt}}^{2}\right) \right]^{1/2}.
\label{eq:vnt}
\end{equation}
Here $\Delta\lambda_{\mathrm{FWHM}}$ is the full width at half maximum (FWHM) of the line, $\Delta\lambda_{\mathrm{Inst}}$ is the instrumental width, $T_{\mathrm i}$ is the ion temperature, $M$ the ion mass, $k_{\mathrm{B}}$ is the Boltzmann constant, $c$ is the speed of light, and $v_{\mathrm{nt}}$ accounts for line broadening from non-thermal fluid flows, such as turbulence. 

	An effective velocity can be defined as
\begin{equation}
v_{\mathrm{eff}} \equiv \sqrt{\left(\frac{2 k_{\mathrm{B}}T_{\mathrm{i}}}{M} + v_{\mathrm{nt}}^{2}\right)}.
\label{eq:defVeff}
\end{equation}
To determine $v_{\mathrm{eff}}$ we analyzed only the lines in Table~\ref{table:linelist} that were not affected by blending. We also do not consider lines at positions where the scattered light contributes greater than 25\% of the total line intensity. The scattered light comes from outside the field of view and can be expected to have different emission line profiles than emission from within the field of view. The widths of the lines were determined by the Gaussian fitting routine described above. We determined for each line a value of $v_{\mathrm{eff}}$ given by 
\begin{equation}
v_{\mathrm{eff}} = \frac{c}{\lambda} \sqrt{ \frac{\Delta\lambda_{\mathrm{FWHM}}^2 - \Delta \lambda_{\mathrm{Inst}}^{2}}{4 \ln(2)} }.
\label{eq:veff}
\end{equation}

	We used the method of \citet{Tu:ApJ:1998} to calculate the ion temperature. This method assumes that all the ions have the same non-thermal velocity. Thus, the maximum value for the non-thermal velocity that can be consistent with the measured width of every ion is the effective velocity $v_{\mathrm{eff}}$ of the narrowest observed line with the assumption that for that ion $T_{\mathrm{i}} = 0$. The lower bound on the ion temperature is obtained by setting $v_{\mathrm{nt}}$ equal to this maximum value in equation~(\ref{eq:defVeff}), and solving for $T_{\mathrm{i}}$. The upper bound can be found by assuming that $v_{\mathrm{nt}}=0$ and again using equation~(\ref{eq:defVeff}) to solve for $T_{\mathrm{i}}$. 
	
	We found the maximum value for $v_{\mathrm{nt}}$ in all the observations was in the range $\approx 20 - 40$~$\mathrm{km\, s^{-1}}$. For polar coronal holes during the 1996-1997 minimum \citet{Tu:ApJ:1998} found the maximum value for $v_{\mathrm{nt}}$ was in the range $33-44$~$\mathrm{km\, s^{-1}}$ and \citet{Landi:ApJ:2009} found a range of $30-50$~$\mathrm{km\, s^{-1}}$. The present measurements are in line with these earlier observations of polar coronal holes. 
	
\begin{figure}
\includegraphics[width=0.9\textwidth]{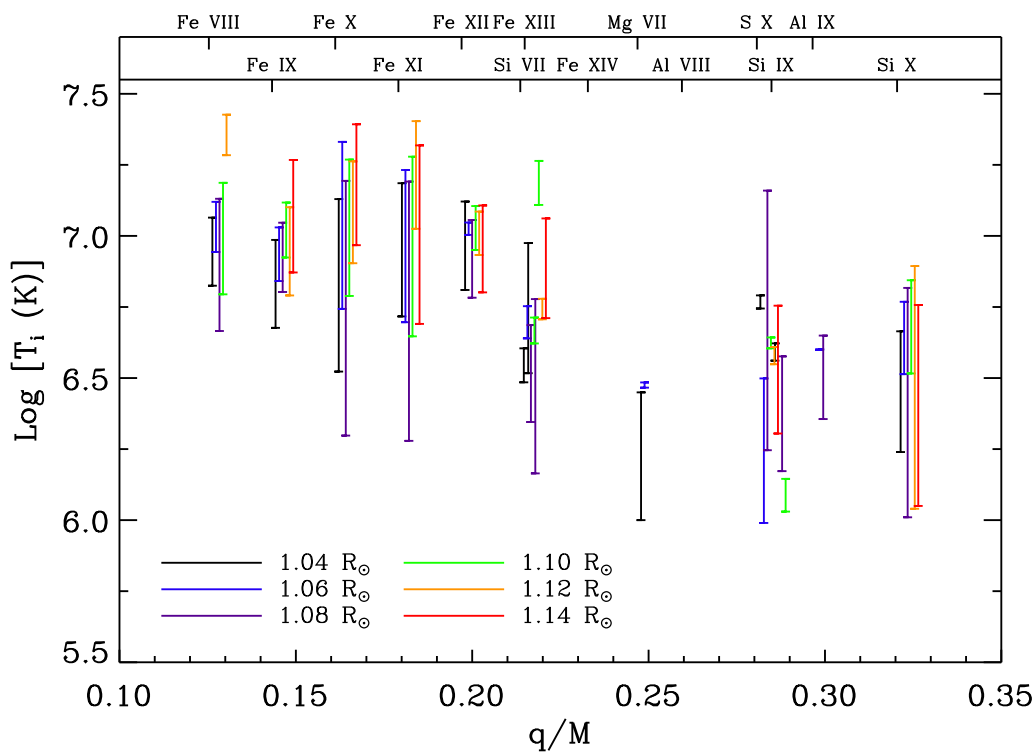}
\caption{\label{fig:Ti_height} Ion temperature range as a function of charge-to-mass ratio for several heights in the $223^{\prime\prime}$ polar coronal hole observation. The behavior for the other pointings was the same, with $T_{\mathrm{i}}$ versus $q/M$ unchanging over the height range of the observations. The data for each ion have been offset slightly in $q/M$ for clarity.}
\end{figure}

\begin{figure}
\includegraphics[width=0.9\textwidth]{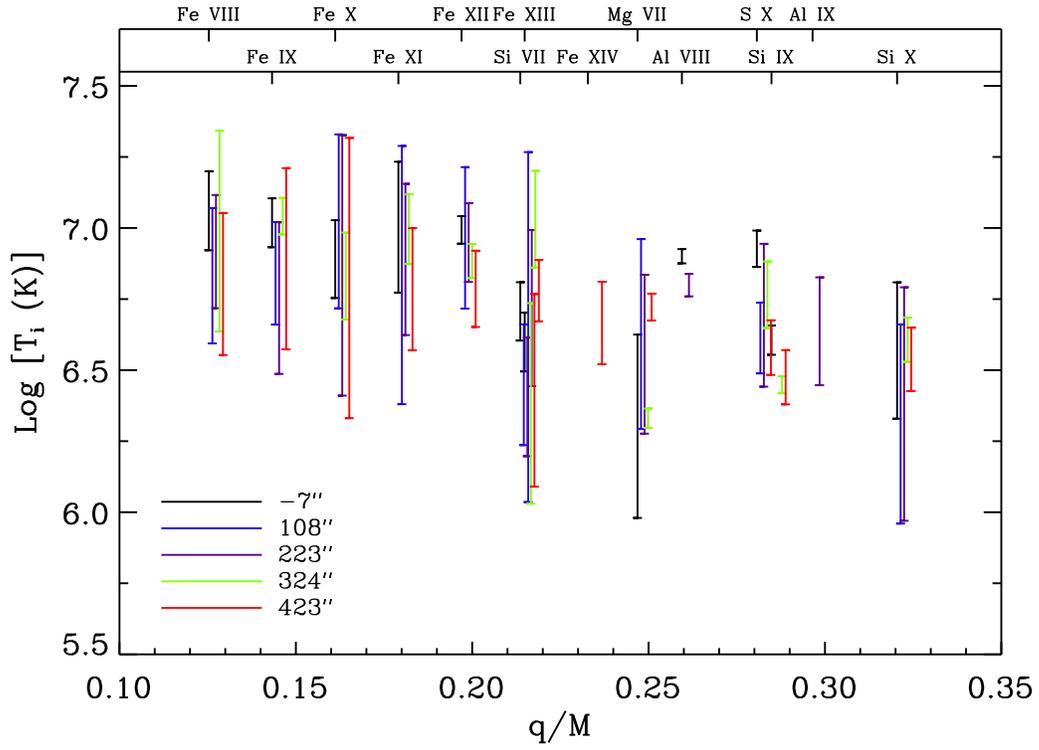}
\caption{\label{fig:Ti_obs} Ion temperature range as a function of charge-to-mass ratio for each observation at a height $1.05$~$R_{\sun}$. The data for each ion have been offset slightly in $q/M$ for clarity.}
\end{figure}

	In all our observations the lower bound on the ion temperature was greater than the electron temperature for all but the narrowest lines where the lower bound was set to zero by definition. Given these results, we adopted a more restrictive lower bound by assuming that for the minimum ion temperature $T_{\mathrm{i}} = T_{\mathrm{e}}$, rather than zero. We then proceeded as before using a corresponding upper limit on $v_{\mathrm{nt}}$. With this assumption the upper bound on $v_{\mathrm{nt}}$ was reduced to $\approx 15-35$~$\mathrm{km\,s^{-1}}$. The ion temperature ranges are plotted as a function of charge-to-mass ratio $q/M$ in Figures~\ref{fig:Ti_height} and \ref{fig:Ti_obs} where $q$ is given in units of $e$ and $M$ in amu. The results are qualitatively the same as under the assumption that the minimum $T_{\mathrm{i}} = 0$, but the tighter bounds on $T_{\mathrm{i}}$ make the dependence on $q/M$ clearer. 
	
	In the observed north polar coronal hole the relation between $T_{\mathrm{i}}$ and $q/M$ is qualitatively the same over the height range of each observation (Figure~\ref{fig:Ti_height}). The temperature of low-charged, high-mass ions is $\log T_{\mathrm{i}} \approx 6.5 - 7.4$, about an order of magnitude greater than the electron temperature. The maximum $T_{\mathrm{i}}$ for $q/m \lesssim 0.2$ ions here is somewhat larger than for the 1996-1997 observations, which found $\log T_{\mathrm{i}} < 7.0$. Both Figures~\ref{fig:Ti_height} and \ref{fig:Ti_obs} show that the ion temperature is lower for $q/M \gtrsim 0.2$. This behavior is similar to that of \citet{Landi:ApJ:2009} for a polar coronal hole during the 1996-1997 solar minimum. \citet{Landi:ApJ:2009} found that $T_{\mathrm{i}}$ increases for $q/M \gtrsim 0.25$, but such behavior was not seen in our analysis. 
	
	For a given height, the behavior of $T_{\mathrm{i}}$ was similar across the observations (Figure~\ref{fig:Ti_obs}). In particular there was no significant difference in the portion of the $423^{\prime\prime}$ observation below $\approx 1.15$~$R_{\sun}$, where we found a higher $T_{\mathrm{e}}$ compared to positions within the polar coronal hole.
		
\begin{figure}
\includegraphics[width=0.9\textwidth]{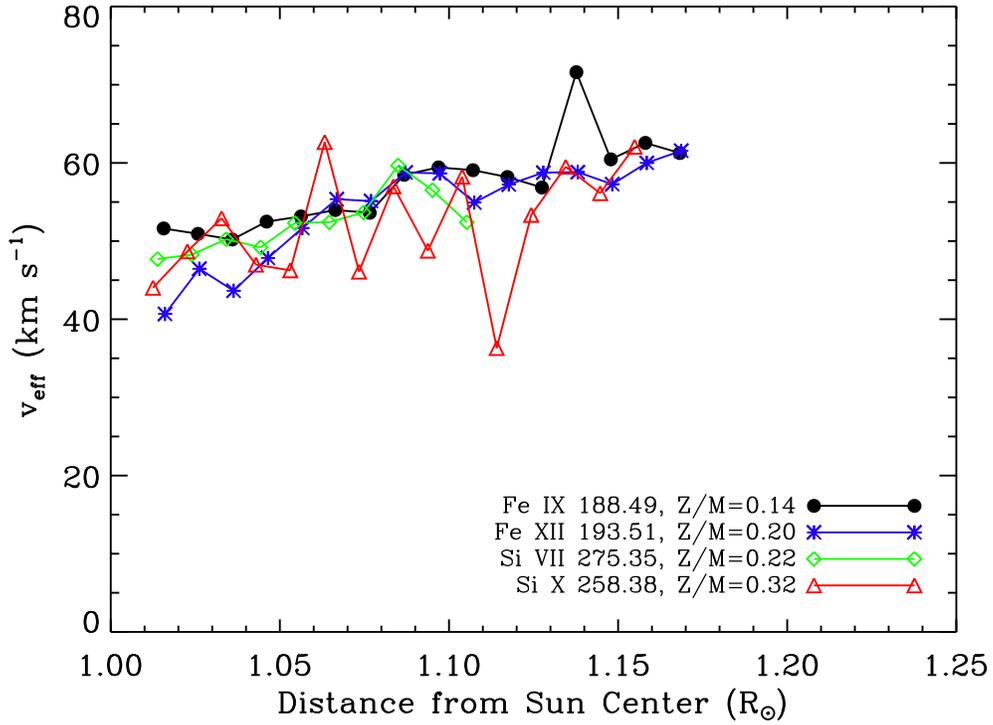}
\caption{\label{fig:veff_vs_height} The effective velocity $v_{\mathrm{eff}}$ in the $223^{\prime\prime}$ observation using equation~(\ref{eq:veff}). In each case the effective velocity increases with height at a similar rate. This could indicate either heating of the ions or increasing non-thermal velocity. The statistical error bars on $v_{\mathrm{eff}}$ are generally smaller than the size of the data points.}
\end{figure}

	Figure~\ref{fig:veff_vs_height} shows how the line widths (i.e. $v_{\mathrm{eff}}$) increase with height, for a selection of lines from ions with different $q/M$. The rate of increase is similar for each of the ions and an increase is found in all observations. There are several potential interpretations for the increasing line width. One possibility is that it reflects the increasing ion temperature with height as suggested in figure~\ref{fig:Ti_height}. This may be due to more heating. The upper bound of $T_{\mathrm{i}}$ increases with height by about $\Delta \log T_{\mathrm{i}} \approx 0.2$ over the range of these observations. The increasing width could also reflect an increasing non-thermal velocity. An increasing $v_{\mathrm{nt}}$ could be caused, for example, by undamped Alfv\'{e}n waves whose velocity amplitude must increase as the density decreases in order to conserve wave energy flux $F$ which is given by
\begin{equation}
F= \sqrt{\frac{\rho}{4\pi}} \avg{\delta v^{2}} B.
\label{eq:Aflux}
\end{equation}	
Here, $B$ is the magnetic field strength, $\rho \approx n_{\mathrm{e}} m_{\mathrm{p}}$ is the mass density with $m_{p}$ the mass of a proton, and $\avg{\delta v^{2}} = 2 v_{\mathrm{nt}}^{2}$ is the mean square velocity perturbation from the wave \citep{Jacques:ApJ:1977, Moran:AA:2001}. Similar increasing line widths were also observed in coronal holes during the previous solar minimum \citep{Banerjee:AA:1998, Wilhelm:ApJ:1998, Doschek:ApJ:2001}. 
		
\section{Discussion and Summary} \label{sec:dis}

	These measurements of a north polar coronal hole at low heights from the recent solar minimum show broad similarities with polar coronal hole observations from the 1996-1997 minimum. The results are summarized in Table~\ref{table:summary}. Similarities in the plasma properties of polar coronal holes between the two minima were also found by \citet{Miralles:ASP:2010} in polar coronal holes above 1.7~$R_{\sun}$ using the \emph{SOHO} Ultraviolet Coronagraph Spectrometer \citep[UVCS;][]{Kohl:SolPhys:1995}. 

		The measured electron density is roughly the same as that measured for polar coronal holes during the previous minimum. \textit{In situ} observations of the solar wind above $1.4$ AU show that the density of the fast solar wind decreased by $\approx 20\%$ compared to the 1996-1997 minimum. The present coronal observation that the density is comparable between the two minima at low heights is not inconsistent with the solar wind measurements since density flux can be conserved through changes in the magnetic field.
	
		The electron temperatures measured using the crossings of the EM curves were in excellent agreement in both magnitude and gradient with height above the solar limb between the 1996-1997 solar minimum measurements and the present results. The $423^{\prime\prime}$ observation showed a transition in $T_{\mathrm{e}}$ from the quiet Sun corona with $\log T_{\mathrm{e}} \approx 6.15$ to the polar coronal hole with $\log T_{\mathrm{e}} \approx 6.03$ near $1.15$~$R_{\sun}$. However, we did not observe any significant transitions for other measured properties at that location. Using the emission measure decay to estimate the scale height in each observation, we found the present observations are closer to hydrostatic equilibrium than was seen for 1996-1997. 
	
	We also compared FIP factors relative to iron and found results for silicon and aluminum similar in the polar coronal hole to what was found in earlier solar minimum polar coronal hole observations. All three elements have low FIPs and are expected to have FIP factors that behave approximately the same way. For the low FIP element magnesium, we found a slightly larger FIP factor relative to iron, but the magnesium analysis is subject to some additional systematic errors. 
	
	For the moderately high FIP element sulfur, we found $f_{\mathrm{S}}/f_{\mathrm{Fe}} \sim 1$ in the polar coronal hole. This measurement for elements with significantly different FIPs suggests that a FIP effect was not present in the polar coronal hole because in the absence of a FIP effect we expect $f_{\mathrm{S}}/f_{\mathrm{Fe}} \approx 1.0$. Previous observations have not found a significant FIP effect in coronal holes. However, \citet{Feldman:ApJ:1998} studied a polar coronal hole during the previous solar minimum and reported a FIP factor for sulfur $f_{\mathrm{S}}/f_{\mathrm{Fe}} < 1$ consistent with the FIP effect. But they found that no other elements of either low or high FIP, except for sulfur, showed any indication of a FIP effect. This may be explained by revisions in the most recent elemental abundances by \citet{Asplund:ARAA:2009} compared to earlier data such as \citet{Feldman:PhysScr:2000}. 
		
	It would be desireable to perform these FIP factor ratio measurements with a larger variety of elements in the near future. We measure ratios of FIP factors relative to iron, because the EIS spectrum had a large number of strong iron lines which provided a robust measurement. It would be better to measure FIP factor ratios by comparing to high FIP elements such such as oxygen, nitrogen, argon, or neon, which are expected to experience only a very weak FIP effect. Unfortunately, lines from these elements were not available. The EIS spectrum should include lines from the oxygen ions O~\textsc{iv}$-$O~\textsc{vi}, but the formation temperature for these ions is less than $\log T_{\mathrm{e}} = 5.5$, well below coronal temperatures. Similarly, argon ions Ar~\textsc{xi} and Ar~\textsc{xiv} may be found in the spectrum; but at $\log T_{\mathrm{e}} = 6.3$ and $6.5$, respectively, the formation temperatures for these ions are much higher than the observed temperature. 
	
	The inferred ion temperature decreases with increasing ion $q/M$ up to the maximum observed $q/M \lesssim 0.32$. For $q/M < 0.25$ the dependence of $T_{\mathrm{i}}$ on $q/M$ is similar to what was observed in the previous solar minimum. Measurements from the 1996-1997 solar minimum showed an upturn in $T_{\mathrm{i}}$ for $q/M > 0.25$, in contrast to the present results where no upturn was observed. In the polar coronal hole we found a maximum ion temperature of $\log T_{\mathrm{i}} \approx 7.4$ for ions with charge to mass ratios $q/M \lesssim 0.2$. The upper bound is slightly higher than in measurements from the previous solar minimum, but the inferred temperature range is large so that the possibility of $T_{\mathrm{i}}$ being comparable in the 1996-1997 and 2007-2009 minima cannot be ruled out. 
	
	In each observation the line widths were found to increase with height. This can be attributed to more ion heating, less cooling, or an increase of the non-thermal velocity. In the present observation the height range was limited by the actual field of view of the detector and by the low intensity and scattered light contribution at high altitudes. It would be interesting to observe the behavior of $T_{\mathrm{i}}$ at higher heights, which requires coronagraphic occultation such as is done by UVCS. 

	It is surprising that the solar minimum observations are so similar given that polar coronal holes are known sources of the fast solar wind and significant changes have been observed in the solar wind power, density, and velocity compared to the previous minimum. Such changes were not observed in our measurements of the lower regions of the north polar coronal hole. 
		
\acknowledgements
	We thank S. R. Cranmer and J. M. Laming for stimulating discussions. MH and DWS were supported in part by the NASA Solar Heliospheric Physics program. The work of EL and MPM is supported by several NASA grants. The work of PB was performed under contract with the Naval Research Laboratory and was funded by NASA.
	
\clearpage
\begin{center}
\begin{longtable}{l c c r @{ -- } l}
\caption{Line List}\label{table:linelist}\\
\hline \hline
Ion & & $\lambda$ (\AA)\tablenotemark{1} & \multicolumn{2}{c}{Transition\tablenotemark{1} } \\
\hline
\endfirsthead
\hline \hline 
Ion & & $\lambda$ (\AA)\tablenotemark{1} & \multicolumn{2}{c}{Transition\tablenotemark{1} } \\
\hline
\endhead
\hline \hline 
\endlastfoot
Mg \textsc{vii} & &276.154 & $2s^2\, 2p^2\, ^{3}P_{0}$ & $2s\, 2p^3\, ^{3}S_{1}$ \\
Mg \textsc{vii}\tablenotemark{2} & &278.404 & $2s^2\, 2p^2\, ^{3}P_{2}$  & $ 2s\, 2p^{3}\, ^{3}S_{1}$ \\
Mg \textsc{vii} & &280.742 & $2s^2\, 2p^2\, ^{1}D_{2} $ & $ 2s\, 2p^3\, ^{1}P_{1}$ \\
Al \textsc{viii} & &250.139 & $2p^2\, ^{3}P_{2}$ & $2s\, 2p^{3}\, ^{3}S_{1}$ \\
Al \textsc{ix} & &280.135 & $2s^2\, 2p\, ^{2}P_{1/2}$ & $2s\, 2p^2\, ^{2}P_{3/2}$ \\
Al \textsc{ix} & &284.025 & $2s^2\, 2p\, ^{2}P_{3/2}$ & $2s\, 2p^2\, ^{2}P_{3/2}$ \\
Al \textsc{ix} & &286.376 & $2s^2\, 2p\, ^{2}P_{3/2}$ & $2s\, 2p^2\, ^{2}P_{1/2}$ \\
Si \textsc{vii} & &272.648 & $ 2s^2\, 2p^4\, ^{3}P_2 $ & $2s\, 2p^5\, ^{3}P_1 $ \\
Si \textsc{vii} & &275.361 & $ 2s^2\, 2p^4\, ^{3}P_2 $  & $2s\, 2p^5\, ^{3}P_2 $ \\
Si \textsc{vii} & &275.676 & $ 2s^2\, 2p^5\, ^{3}P_1 $  & $2s\, 2p^5\, ^{3}P_1 $ \\
Si \textsc{ix} & &258.082 & $ 2s^2\, 2p^2\, ^{1}D_2 $  & $2s\, 2p^3\, ^{1}D_2 $\\
Si \textsc{x} & &258.371 & $ 2s^2\, 2p\, ^2P_{3/2} $ & $2s\, 2p^2\, ^{2}P_{3/2} $ \\
Si \textsc{x} & &261.044 & $2s^2\, 2p\, ^2P_{3/2} $ & $2s\, 2p^2\, ^{2}P_{1/2} $ \\
Si \textsc{x} & &272.006 & $ 2s^2\, 2p\, ^2P_{1/2} $ & $2s\, 2p^2\, ^{2}S_{1/2} $ \\
Si \textsc{x} & &277.278 & $ 2s^2\, 2p\, ^{2}P_{3/2} $ & $ 2s\, 2p^2\, ^{2}S_{1/2} $ \\
S \textsc{viii} \tablenotemark{3} & &198.554 & $2s^2\, 2p^5\, ^2P_{3/2}$ & $2s\, 2p^{6}\, ^{2}S_{1/2}$ \\ 
S \textsc{viii} \tablenotemark{4} & &202.610 & $2s^2\, 2p^5\, ^2P_{1/2}$ & $2s\, 2p^{6}\, ^{2}S_{1/2}$ \\
S \textsc{x} & &257.147 & $2s^2\, 2p^3\, ^{4}S_{3/2}$  & $2s\, 2p^{4}\, ^{4}P_{1/2}$ \\
S \textsc{x} & &264.231 & $2s^2\, 2p^3\, ^{4}S_{3/2}$ & $2s\, 2p^{4}\, ^{4}P_{5/2}$ \\
Fe \textsc{viii} & &185.213 & $ 3p^6\, 3d\, ^{2}D_{5/2} $ & $3p^5\,3d^2\, (^{3}F)\, ^{2}F_{7/2}$ \\
Fe \textsc{viii} & & 186.599 & $3p^6 \,3d\, ^{2}D_{3/2} $ & $3p^5\, 3d^2\, (^{3}F)\, ^{2}F_{5/2}$ \\
Fe \textsc{viii} & & 194.661 & $ 3p^6 \, 3d\, ^{2}D_{5/2} $ & $3p^6\, 4p\, ^{2}P_{3/2}$ \\
Fe \textsc{viii} & & 195.972 & $ 3p^6\, 3d\, ^{2}D_{3/2} $ & $3p^6\, 4p\, ^{2}P_{1/2} $ \\
Fe \textsc{viii} & &197.362 & $3p^6\, 3d\, ^{2}D_{5/2}$ & $3p^5\, 3d^2\, (^{1}S)\, ^{2}P_{3/2}$ \\
Fe \textsc{ix} & &188.497 & $ 3s^2\, 3p^5\, 3d\, ^{3}F_4 $ & $3s^2\, 3p^4\, (^{3}P)\, 3d^2\, ^{3}G_5 $\\
Fe \textsc{ix} & &189.941 & $ 3s^2\, 3p^5\, 3d\, ^{3}F_3 $ & $3s^2\, 3p^4\, (^{3}P)\, 3d^2\, ^{3}G_4 $\\
Fe \textsc{ix} & &197.862 & $ 3s^2\, 3p^5\, 3d\, ^{1}P_1 $ & $3s^2\, 3p^5\, 4p\, ^{1}S_0 $ \\
Fe \textsc{x}  & &182.307 & $ 3s^2\, 3p^5\, ^{2}P_{1/2}$  & $3s^2\, 3p^4\, (^{3}P)\, 3d\, ^{2}P_{3/2} $\\
Fe \textsc{x} & &184.537 & $ 3s^2\, 3p^5\, ^{2}P_{3/2} $ & $3s^2\, 3p^4\, 	(^1D)\, 3d\, ^{2}S_{1/2} $ \\
Fe \textsc{x} & &190.037 & $ 3s^2\, 3p^5\, ^{2}P_{1/2} $ & $3s^2\, 3p^4\, (^{1}D)\, 3d\, ^{2}S_{1/2} $ \\
Fe \textsc{x} & &193.715 & $ 3s^2\, 3p^5\, ^{2}P_{3/2} $ & $3s^2\, 3p^4\, (^{1}S)\, 3d\, ^{2}D_{5/2} $ \\
															&	& 257.259 & $ 3s^2\, 3p^5\, ^{2}P_{3/2} $ & $3s^2\, 3p^4\, (^{3}P)\, 3d\, ^{4}D_{5/2} $ \\*
\raisebox{2.5ex}[0pt]{Fe \textsc{x}}&\raisebox{2.5ex}[0pt]{$\Big\{$} & 257.263 & $3s^2\, 3p^5\, ^{2}P_{3/2}$ & $3s^2\, 3p^4\, (^{3}P)\, 3d\, ^{4}D_{7/2}$\\
Fe \textsc{xi} & &180.408 & $ 3s^2\, 3p^4\, ^{3}P_2 $ & $3s^2\, 3p^3\, (^{4}S)\, 3d\, ^{3}D_3 $\\
Fe \textsc{xi} & &182.169 & $ 3s^2\, 3p^4\, ^{3}P_1 $ & $3s^2\, 3p^3\, (^{4}S)\, 3d\, ^{3}D_2 $ \\
Fe \textsc{xi} & &188.232 & $ 3s^2\, 3p^4\, ^{3}P_2 $ & $3s^2\, 3p^3\, ( ^{2}D)\, 3d\, ^{3}P_2 $ \\
Fe \textsc{xi}\tablenotemark{5} & &188.299 & $ 3s^2\, 3p^4\, ^{3}P_2 $ & $3s^2\, 3p^3\, (^{2}D)\, 3d\, ^{1}P_1 $ \\
Fe \textsc{xi} & &189.719 & $3s^2\, 3p^4\, ^{3}P_{0}$ & $3s^2\, 3p^3\, (^{2}D)\, 3d\, ^{3}P_{1}$ \\
Fe \textsc{xii} & &192.394 & $3s^2\, 3p^3\, ^{4}S_{3/2}$ & $3s^2\, 3p^2\, (^{3}P)\, 3d\, ^{4}P_{1/2}$ \\
Fe \textsc{xii} & &193.509 & $3s^2\, 3p^3\, ^{4}S_{3/2}$ & $3s^2\, 3p^2\, (^{3}P)\, 3d\, ^{4}P_{3/2}$ \\
 			& &195.119 & $ 3s^2\, 3p^3\, ^{4}S_{3/2} $ & $3s^2\, 3p^2\, (^{3}P)\, 3d\, ^{4}P_{5/2} $ \\*
\raisebox{2.5ex}[0pt]{Fe \textsc{xii}} &\raisebox{2.5ex}[0pt]{$\Big\{$} &195.179 & $ 3s^2\, 3p^3\, ^{2}D_{3/2} $ & $3s^2\, 3p^2\, (^{1}D)\, 3d\, ^{2}D_{3/2} $ \\
Fe \textsc{xiii} & &202.044 & $ 3s^2\, 3p^2\, ^{3}P_0 $ & $3s^2\, 3p\, 3d\, ^{3}P_1 $ \\
      & &203.797 & $ 3s^2\, 3p^2\, ^{3}P_{2}$ & $ 3s^2\, 3p\, 3d\, ^{3}D_{2}$ \\*
\raisebox{2.5ex}[0pt]{Fe \textsc{xiii}}	&\raisebox{2.5ex}[0pt]{$\Big\{$} &203.828 & $ 3s^2\, 3p^2\, ^{3}P_{2}$ & $ 3s^2\, 3p\, 3d\, ^{3}D_{3}$
\footnotetext[1]{Wavelengths and transitions given by \textsc{chianti} \citep{Dere:AA:2009}.}
\footnotetext[2]{Si~\textsc{vii} $\lambda 278.44$ blend subtracted as described in Section~\ref{sec:anal}.}
\footnotetext[3]{Fe~\textsc{xi} $\lambda 198.55$ blend subtracted as described in Section~\ref{sec:anal}.}
\footnotetext[4]{Fe~\textsc{xi} $\lambda 202.63$  blend subtracted as described in Section~\ref{sec:anal}.}
\footnotetext[5]{\citet{Brown:ApJS:2008} give the upper state as $^{3}S_{1}$.}
\renewcommand{\thefootnote}{\fnsymbol{footnote}}
\footnotetext[0]{None of the lines where blends were subtracted (annotated 1 -- 4) nor the self-blended lines (marked by a curly bracket) were used in the ion temperature analysis.}
\end{longtable}
\end{center}
\renewcommand{\thefootnote}{\alph{footnote}}

\clearpage

\begin{center}
\begin{longtable}{lcc}
\caption{Electron temperatures based on the scale height of $\log \mathrm{EM}(R)$.}\label{table:Tscale}\\
\hline \hline 
Observation & $\log T_{\mathrm{e}}$ Fe lines & $\log T$ Si lines  \\
\hline
\endfirsthead
\hline \hline 
Observation & $\log T_{\mathrm{e}}$ Fe lines & $\log T$ Si lines  \\
\hline
\endhead
\hline\hline
\endlastfoot
$-7^{\prime\prime}$ & $6.24 \pm 0.14 $ & $6.08 \pm 0.03$\\
108$^{\prime\prime}$ & $6.23 \pm 0.12$ & $6.16 \pm 0.06$ \\
223$^{\prime\prime}$ & $6.17 \pm 0.08$ & $6.08 \pm 0.02$ \\
324$^{\prime\prime}$ & $6.10 \pm 0.07$ & $ 6.01 \pm 0.02$ \\
423$^{\prime\prime}$ & $6.04 \pm 0.04$ & $ 6.00 \pm 0.01 $ \\
\end{longtable}
\end{center}

\clearpage
\begin{center}
\begin{longtable}{lcccccc} 
\caption{Weighted mean FIP factor ratios.}\label{table:fipratio}\\
\hline\hline
Ratio & \multicolumn{6}{c}{Observation} \\
 & $-7^{\prime\prime}$ & $108^{\prime\prime}$ & $223^{\prime\prime}$ & $324^{\prime\prime}$ & \multicolumn{2}{c}{$423^{\prime\prime}$} \\
  & (CH) & (CH) & (CH) & (CH) & $R > 1.15$~$R_{\sun}$ (CH) & $R < 1.15$~$R_{\sun}$ (QS) \\
\hline
\endfirsthead

\endhead
\hline\hline
\endlastfoot
$\avg{f_{\mathrm{Si}}/f_{\mathrm{Fe}}}$ & $1.15_{-0.19}^{+0.21}$ & $1.10_{-0.17}^{+0.21}$ & $1.22_{-0.18}^{+0.21}$ & $1.24_{-0.16}^{+0.17}$ & $1.45_{-0.33}^{+0.44}$ & $1.42_{-0.19}^{+0.22}$ \\
$\avg{f_{\mathrm{S}}/f_{\mathrm{Fe}}}$ & $1.17_{-0.19}^{+0.22}$ & $1.04_{-0.16}^{+0.19}$ & $1.17_{-0.18}^{+0.21}$ & $1.03_{-0.12}^{+0.15}$ & $0.82_{-0.17}^{+0.23}$ & $1.12_{-0.15}^{+0.18}$\\
$\avg{f_{\mathrm{Al}}/f_{\mathrm{Fe}}}$ & $1.38_{-0.47}^{+0.71}$ & $1.66_{-0.60}^{+0.95}$ & $1.51_{-0.30}^{+0.39}$ & 
$1.19_{-0.22}^{+0.29}$ & \nodata & $1.43_{-0.29}^{+0.38}$ \\
$\avg{f_{\mathrm{Mg}}/f_{\mathrm{Fe}}}$ & $1.32_{-0.27}^{+0.34}$ & $1.57_{-0.30}^{+0.38}$ & $1.41_{-0.27}^{+0.32}$ & $2.44_{-0.58}^{+0.76}$ & $1.88_{-0.75}^{+1.25}$ & \nodata \\ 
\renewcommand{\thefootnote}{\fnsymbol{footnote}}
\footnotetext[0]{(CH) - The measurement lies in the polar coronal hole. (QS) - The high $T_{\mathrm{e}}$ at this location suggests the measurement may be in the quiet Sun corona.}
\end{longtable}
\end{center}
\renewcommand{\thefootnote}{\alph{footnote}}

\clearpage

\scriptsize
\begin{center}
\begin{longtable}{lcccl}
\caption{Observed polar coronal hole properties compared to some representative values for the 1996-1997 solar minimum.}\label{table:summary} \\
\hline\hline
Quantity & 2007\tablenotemark{a} & 1996--1997 & Reference\tablenotemark{b} \\
\hline
\endfirsthead

\endhead
\hline\hline
\endlastfoot
$n_{\mathrm{e}} (1.05$~$R_{\sun})$	& $\approx 8 \times 10^{7}$~$\mathrm{cm^{-3}}$ & $\approx 7 \times 10^{7}$~$\mathrm{cm^{-3}}$ & \citet{Wilhelm:ApJ:1998} \\
GEM $\log T_{\mathrm{e}} (1.05$~$R_{\sun})$ 	& $5.97 \pm 0.02$\tablenotemark{c}		&	$5.95 \pm 0.04$ 	& \citet{Landi:ApJ:2008} \\
$\Delta \log T_{\mathrm{e}} / \Delta R$	& $\approx 0.75$~$R_{\sun}^{-1}$ & $\approx 0.73$~$R_{\sun}^{-1}$ & \citet{Landi:ApJ:2008}\\
Scale Height $\log T_{\mathrm{e}}$	& 6.05 --  6.38 & 6.54 & \citet{Landi:ApJ:2008} \\
$f_{\mathrm{Si}}/f_{\mathrm{Fe}}$ 	& $1.15^{+0.21}_{-0.19}$\tablenotemark{d} & $\approx 1$ & \citet{Feldman:ApJ:1998} \\
$f_{\mathrm{S}}/f_{\mathrm{Fe}}$ 	& $1.17^{+0.22}_{-0.19}$\tablenotemark{d} & $\approx 0.7$ & \citet{Feldman:ApJ:1998} \\
$f_{\mathrm{Al}}/f_{\mathrm{Fe}}$	& $1.38^{+0.71}_{-0.47}$\tablenotemark{d} & \nodata & \nodata \\
$f_{\mathrm{Mg}}/f_{\mathrm{Fe}}$	& $1.32^{+0.34}_{-0.27}$\tablenotemark{d} & $\approx 1$ & \citet{Feldman:ApJ:1998}\\
$v_{\mathrm{nt}}$ 									& $ \lesssim 40$~$\mathrm{km\,s^{-1}}$ & $\lesssim 50$~$\mathrm{km\,s^{-1}}$ & \citet{Landi:ApJ:2009} \\
$\log T_{\mathrm{i}}(q/M < 0.2)$ 	& 6.5 -- 7.4 & 6.3 -- 7.0 & \citet{Landi:ApJ:2009}\\
$T_{\mathrm{i}}$ vs. $q/M$ 				& decreasing & decreasing $q/M < 0.25$; increasing $q/M > 0.25$ & \citet{Landi:ApJ:2009} \\
$\Delta v_{\mathrm{eff}}/\Delta R$ & $\approx 70$~$\mathrm{km\,s^{-1}\,R_{\sun}^{-1}}$	& 40 -- 70~$\mathrm{km\,s^{-1}\,R_{\sun}^{-1}}$	& \citet{Wilhelm:ApJ:1998} \\
\footnotetext[1]{This paper}
\footnotetext[2]{1996--1997 minimum}
\footnotetext[3]{Based on GEM analysis of silicon lines.}
\footnotetext[4]{From the $-7^{\prime\prime}$ observation.}
\end{longtable}
\end{center}


\clearpage

\clearpage

\begin{figure}
\includegraphics[width=0.9\textwidth]{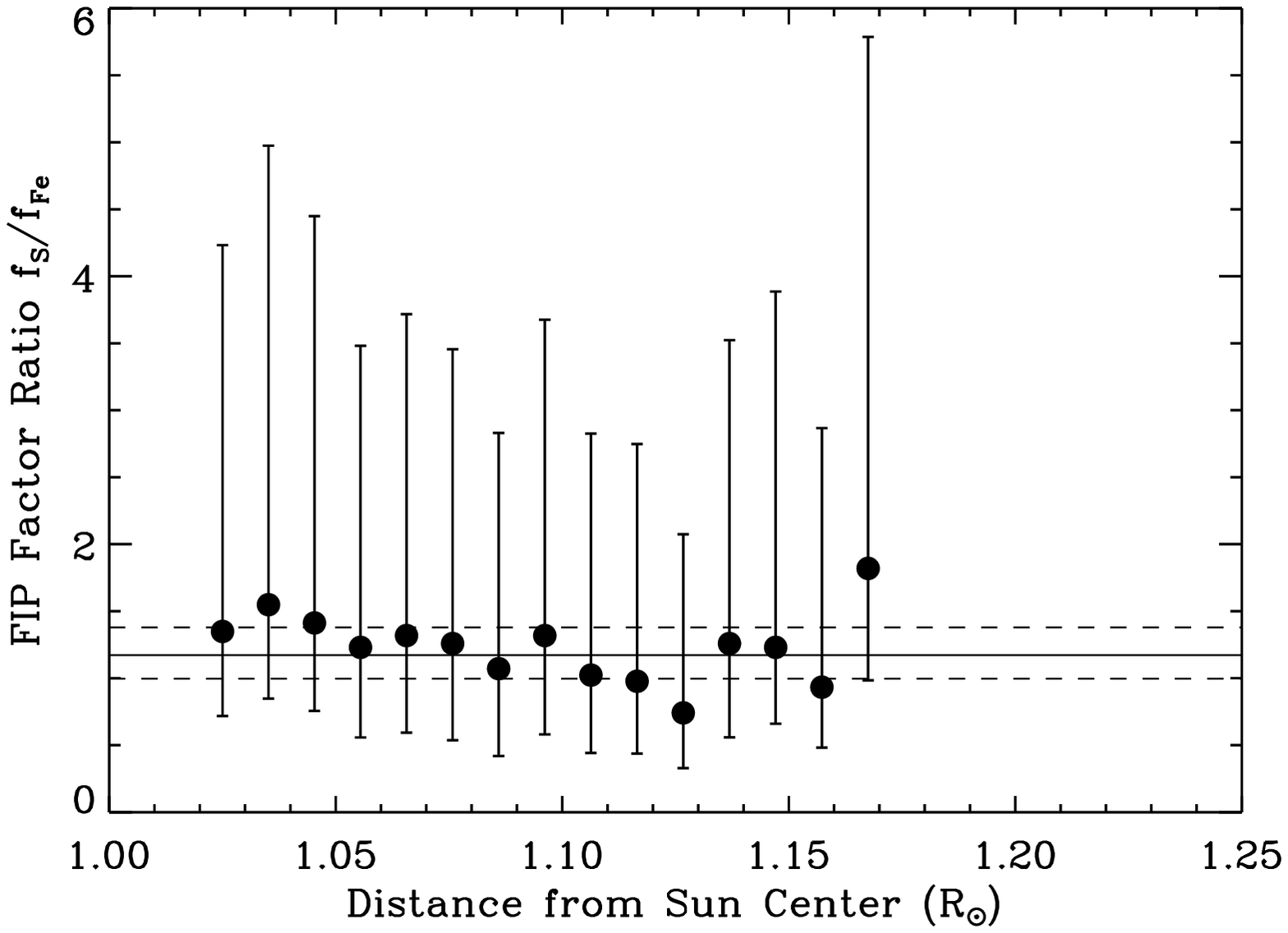}
\caption{\label{fig:Sratio} Same as Figure~\ref{fig:EMratio}, but for $f_{\mathrm{S}}/f_{\mathrm{Fe}}$.}
\end{figure}

\bibliography{Solar_bib}

\end{document}